\pdfoutput=1 

\documentclass[usenatbib,useAMS,pdftex]{mn2e}

\usepackage{times}
\usepackage{rotating}
\usepackage{graphicx}
\usepackage{epstopdf}
\usepackage{bbold}
\usepackage{multirow}
\usepackage{url}

\usepackage{amsmath} 
\usepackage{amssymb}    

\usepackage[rightcaption]{sidecap}

\usepackage{xcolor}

\usepackage{float}

\newcommand{\lsim}{\mbox{${\,\hbox{\hbox{$ < $}\kern -0.8em \lower 1.0ex\hbox{$\sim$}}\,}$}}
\newcommand{\gsim}{\mbox{${\,\hbox{\hbox{$ > $}\kern -0.8em \lower 1.0ex\hbox{$\sim$}}\,}$}}

\newcommand{\dd}{{\rm d}}

\newcommand{\cc}{{\rm c}}

\def\beqn{\vspace{2mm}
\begin{eqnarray}} 
\def\eeqn{\vspace{2mm} 
\end{eqnarray}}

 \def\la{\langle}

\newcommand{\be}{\begin{equation}}
\newcommand{\ee}{\end{equation}}
\newcommand{\ba}{\begin{eqnarray}}
\newcommand{\ea}{\end{eqnarray}}
\newcommand{\brr}{\begin{array}}
 
\newcommand{\err}{\end{array}}
\newcommand{\bc}{\begin{center}}
\newcommand{\ec}{\end{center}}

\begin{document}


\title[Constraining the halo bispectrum]{Constraining the halo bispectrum  in real and redshift space \\from   perturbation theory and non-linear stochastic bias}

\author[F.~S.~Kitaura et.~al.]{Francisco-Shu Kitaura$^{1}$\thanks{E-mail: kitaura@aip.de, Karl-Schwarzschild-fellow}, H{\'e}ctor Gil-Mar{\'i}n$^{2}$, Claudia Scoccola$^{3,4,5,6}$, Chia-Hsun Chuang$^{5,6}$,  \and  Volker M{\"u}ller$^{1}$,   Gustavo Yepes$^{5}$ \& Francisco Prada$^{6,7,8}$ \\
$^{1}$Leibniz-Institut f\"ur Astrophysik Potsdam (AIP), An der Sternwarte 16, D-14482 Potsdam, Germany\\
$^{2}$Institute of Cosmology \& Gravitation, University of Portsmouth, Dennis Sciama Building, Portsmouth PO1 3FX, UK\\
$^{3}$Instituto de Astrof{\'\i}sica de Canarias (IAC), C/V{\'\i}a L\'actea, s/n, E-38200 La Laguna, Tenerife, Spain\\
$^{4}$Departamento de Astrof{\'\i}sica, Universidad de La Laguna (ULL), E-38206 La Laguna, Tenerife, Spain\\
$^{5}$Departamento de F{\'i}sica Te{\'o}rica,  Universidad Aut{\'o}noma de Madrid, Cantoblanco, 28049, Madrid, Spain\\
$^{6}$Instituto de F{\'i}sica Te{\'o}rica, (UAM/CSIC), Universidad Aut{\'o}noma de Madrid, Cantoblanco, E-28049 Madrid, Spain\\
$^{7}$Campus of International Excellence UAM+CSIC, Cantoblanco, E-28049 Madrid, Spain \\
$^{8}$Instituto de Astrof{\'i}sica de Andaluc{\'i}a (CSIC), Glorieta de la Astronom{\'i}a, E-18080 Granada, Spain\\
}

\maketitle

\begin{abstract}

We present a method to produce mock galaxy catalogues with efficient perturbation theory schemes, which match the number density, power spectra and bispectra in real and in redshift space from N-body simulations. The essential contribution of this work is the way in which we constrain the bias parameters in the  \textsc{patchy}-code. In addition of aiming at reproducing the two-point statistics, we seek the set of bias parameters, which constrain the univariate halo probability distribution function (PDF) encoding higher-order correlation functions. We demonstrate that halo catalogues based on the same underlying dark matter field with a fix halo number density, and accurately matching the power spectrum (within 2\%), can lead to very different bispectra depending on the adopted halo bias model. A model ignoring the shape of the halo PDF can lead to deviations up to factors of 2. The catalogues obtained additionally constraining the shape of the halo PDF can significantly lower the discrepancy in the three-point statistics, yielding closely unbiased bispectra both in real and in redshift space; which are in general compatible with those corresponding to an N-body simulation within 10\% (deviating at most up to 20\%). Our calculations show that the constant linear bias of $\sim$2 for Luminous Red Galaxy (LRG) like galaxies seen in the power spectrum, mainly comes from sampling halos in high density peaks, choosing a high density threshold rather than from a factor multiplying the dark matter density field. Our method contributes towards an efficient modelling of the halo/galaxy distribution required to estimate uncertainties in the clustering measurements from galaxy redshift surveys. We have also demonstrated that it represents a powerful tool to test various bias models.

\end{abstract}


\section{Introduction}

Mock galaxy catalogues have become an essential tool to assess systematics in the interpretation of galaxy surveys, to test and develop large-scale structure analysis tools, and to understand structure and galaxy formation. 

However, the new generation of galaxy surveys with increasing volumes and number densities, such as  WiggleZ\footnote{\url{http://wigglez.swin.edu.au/site/}} \citep[][]{wigglez2010}, VIPERS\footnote{\url{http://vipers.inaf.it/}} \citep[][]{guzzo2013}, BOSS\footnote{\url{http://www.sdss3.org/surveys/boss.php}} \citep[][]{boss2011}, DESI\footnote{\url{http://desi.lbl.gov/}}/BigBOSS \citep[][]{bigboss2011}, DES\footnote{\url{http://www.darkenergysurvey.org}} \citep[][]{des2013}, LSST \footnote{\url{http://www.lsst.org/lsst/}} \citep[][]{lsst2012}, J-PAS\footnote{\url{http://j-pas.org/}} \citep[][]{jpas2014}, 4MOST\footnote{\url{http://www.aip.de/en/research/research-area-ea/research-groups-and-projects/4most}} \citep[][]{4most} or Euclid\footnote{\url{http://www.euclid-ec.org}} \citep[][]{2009ExA....23...39C,euclid2009}, require  challenging computational resources to produce the corresponding mock galaxy catalogues necessary to accurately assess the uncertainties in the measurements.  Some remarkable attempts producing large $N$-body simulations can be found in the recent literature \citep[see e.g.][]{KimHorizon2009,AnguloXXL2012,Prada2012,DeusSimulation2012,jubilee2013}.
Nevertheless, these simulations  do not go beyond the high mass end of haloes modelling luminous red galaxies (LRGs). Moreover, they provide a limited number of realizations, which cannot be used to make a proper study of the cosmic variance because the estimation of covariance matrices requires a large number of mock catalogues \citep[][]{2014MNRAS.439.2531P}. 

Interesting alternatives to produce large numbers of $N$-body simulations have been recently proposed, such as re-scaling $N$-body simulations to account for a change in the cosmological parameters  \citep[][]{2010MNRAS.405..143A}, computing covariance matrices from a small set of  simulations \citep[][]{2011ApJ...737...11S}, or including Lagrangian perturbation theory (LPT) within the Vlasov equations solver to speed up  $N$-body codes \citep[\textsc{Cola},][]{cola2013,2014AAS...22345717K}. 

As an alternative to run $N$-body cosmological simulations, one can calibrate approximate structure formation models to $N$-body solutions and scan the parameter space using  more efficient schemes, as the one presented in this work.

  A number of approaches has been proposed in the literature for the generation of mock galaxy catalogues based on LPT, such as \textsc{Pinocchio} \citep[][]{2002ApJ...564....8M,monaco2013}, \textsc{PThalos} \citep[][]{scocci,manera12}, and \textsc{patchy} \citep[][]{patchy}. Alternatively, approximate particle mesh based codes: \textsc{QPM} \citep[][]{2014MNRAS.437.2594W} can be applied.  Another approach consists of producing low resolution $N$-body simulations and augment them with a bias model \citep[][]{delaTorre2012,2013arXiv1310.3880A}. We note that the bias model adopted in \textsc{patchy} can be applied for such kind of approaches as well.   It has been shown  that perturbation theory can provide an accurate  approach to model the Baryon Acoustic Oscillations (BAOs) \citep[][]{2012JCAP...04..013T}. 

In the era of precision cosmology we need to produce mock catalogues that yield the right matter statistics to high accuracy not only in terms of the number density and correlation function (or power spectrum), but also in terms of the higher-order statistics. 
The three-point function is essential for an accurate description of the clustering, as it represents a measure of gravitationally induced non-Gaussianities, which characterises the morphology of the cosmic web \citep{Frieman1994} and has a long history applied to galaxy  surveys \citep[see e.~g.~][]{Gaztanaga1994,Scoccimarro2001,Verde2002,Jing2004,Mueller2004,Mueller2011}. It can be used  to test gravity \citep{Shiratra_et_al_2007,Gil-Marin_et_al_2011}, to break degeneracies in the galaxy bias \citep[][Gil-Mar{\'i}n et al., in prep.]{Matarrese_et_al_1997,Verde_et_al_1998,Scoccimarro_et_al_2001,Verde_et_al_2002,Hoffmann2014}, or to test the existence of primordial non-Gaussianities \citep{Sefusatti_Komatsu_2007,Jeong_Komatsu_2009}.  We note that recent efforts have been performed to constrain the dark matter three-point statistics based on an effective field theory description of the large-scale structure \citep[][]{Angulo2014,Baldauf2014}. An efficient method based on perturbation theory able to produce halo catalogues with the right three-point statistics is still missing.
 
The distribution of haloes is statistically determined by all its moments.
Nevertheless, a method imposing all the corresponding correlation functions (assuming that they are known) to a distribution of haloes is far from trivial and hardly numerically efficient \citep[see][]{kitaura_skewlog}.
Instead, one tries to encode the physics encapturing all the higher-order statistics in the generation of the halo distribution. This is naturally given in a $N$-body simulation \citep[although the specific halo-finder can lead to some discrepancies, see][]{2011MNRAS.415.2293K}.
However, when trying to produce thousands of mock halo catalogues on huge volumes,  $N$-body simulations turn out to be computationally very expensive,  as they require to solve the interaction between dark matter particles on small scales and high resolutions to resolve the different populations of haloes.

 As an alternative one can obtain the large-scale structure cosmic density field (at scales larger than the diameter of the largest haloes, i.e., larger than a few Mpc) with approximate gravity solvers (or from low resolution $N$-body simulations) and model its relation to the halo distribution with a parametrised  statistical bias  description to produce the full halo catalogue or augment the missing populations of haloes. The advantage of such an approach is that it is much faster and less memory consuming, as one does not require high resolutions since the individual haloes do not need to be resolved. Moreover, one can get insights into the halo bias and find accurate descriptions, useful for the analysis of the large-scale structure.

This paper is structured as follows: in the next section (\S \ref{sec:method}) we present our method. We then show  (\S \ref{sec:results}) our numerical experiments calibrating our mock catalogues with $N$-body simulations. Finally (\S \ref{sec:conc}) we present our conclusions and discussion.
 
\section{Method}

\label{sec:method}

Let us start defining the generation of a halo distribution as a statistical problem.
\subsection{Statistical problem}
\label{sec:statsprob}

The distribution of haloes is statistically determined by its mean number density ($\xi^h_1$: one-point correlation function), ($\xi^h_2$: two-point) correlation function, skewness ($\xi^h_3$: three-point correlation function), kurtosis  ($\xi^h_4$: four-point correlation function), and all the rest of higher-order correlation functions:
\begin{equation}
\label{eq:samp1}
N_h \curvearrowleft {\mathcal P}(N_h|\xi^h_1,\xi^h_2,\xi^h_3,\xi^h_4,\dots) {,}
\end{equation} 
where $N_h$ are the number counts of haloes per cosmic sub-volume (per cells when dividing the cosmic volume in a grid), $\xi^h_i$ the different correlation functions corresponding to the halo distribution and  ${\mathcal P}$ the corresponding probability distribution function (PDF).
The halo population is defined in a particular mass range ($N_h=N_h[\Delta M_h]$) or alternatively in a maximum circular velocity range ($N_h=N_h[\Delta v_{\rm max}]$)  and consequently all correlation functions also depend on that range ($\xi^h_i=\xi^h_i[\Delta M_h]$ or $\xi^h_i=\xi^h_i[\Delta v_{\rm max}]$,  for $i=1,\dots,m$ until the highest considered order $m$).

Let us suppose that the expected continuous halo density field $\rho_h=\langle N_h\rangle$ and the PDF relating it with the discrete halo number count per cell are known.
Then we could get the discrete halo distribution by sampling from this PDF:
\begin{equation}
N_h\curvearrowleft {\mathcal P}(N_h|\rho_h) {.}
\end{equation} 
This PDF deviates in general from Poissonity, due to the non-vanishing correlation of haloes producing over- or under-dispersed distributions depending on the halo population and density regime \citep[see][]{2001MNRAS.320..289S,2002MNRAS.333..730C,2014MNRAS.441..646N}.
  Under-dispersion is a sub-dominant effect, which appears when the correlation function becomes negative \citep[see also][]{2012PhRvD..86h3540B,baldauf2013}. Therefore we focus on modelling  over-dispersion with the negative binomial distribution function, which requires a single stochastic bias parameter $\beta$ \citep[see][]{patchy,2014MNRAS.441..646N}.
Stochasticity in the bias relation has since long been studied \citep[see e.~g.~][]{1974ApJ...187..425P,1985MNRAS.217..805P,bbks1986,1993ApJ...413..447F,1996MNRAS.282..347M,1999ApJ...520...24D,1999MNRAS.304..767S,2000MNRAS.318..203S,2002ApJ...575..587B,2007PhRvD..75f3512S,2010PhRvD..82j3529D,2011PhRvD..83j3509B,2011A&A...527A..87V,2012MNRAS.421.3472E,2012PhRvD..85h3509C,2012PhRvD..86h3540B,baldauf2013} and has a non-trivial relation to the non-local bias (see discussion at the end of next section).

To obtain the  expected halo density field  one needs to control the bias of the particular halo population, which in general is a non-linear, scale-dependent, and non-local function of the dark matter density field $\rho_{\rm M}$: $B(\rho_h|\rho_{\rm M})$. 

From a statistical perspective, this bias relation is not only a function of the underlying dark matter field, but also of all the moments of the halo distribution: $B(\rho_h|\rho_{\rm M})=B(\rho_h|\rho_{\rm M},\xi^h_1,\xi^h_2,\xi^h_3,\xi^h_4,\dots)$.
Hence, the haloes are sampled from the following PDF:
\begin{equation}
\label{eq:samp2}
N_h\curvearrowleft{\mathcal P}(N_h|B(\rho_h|\rho_{\rm M},\xi^h_1,\xi^h_2,\xi^h_3,\xi^h_4,\dots)) {.}
\end{equation} 
There is no apparent advantage in the last expression with respect to Eq.~\ref{eq:samp1}, as we still need all the moments of the halo distribution and in addition the dark matter field together with the halo bias. However, there is a striking difference, since we have extended the purely statistical problem with a physical model. The physics is encoded in the large-scale structure density field dominated by dark matter and its biased relation to the halo population.
While the relation in Eq.~\ref{eq:samp1} does not tell us how to obtain the halo distribution \citep[for some attempts based on expansions of Gaussian/Lognormal fields including higher-order statistics see][]{kitaura_skewlog}, Eq.~\ref{eq:samp2} gives us a straightforward approach. In principle we just need to define a parametrised bias model and constrain the corresponding parameters with the halo correlation functions extracted from $N$-body simulations.

\subsection{Deterministic bias relations}

Let us define the parametrised deterministic bias relation.
\begin{enumerate}
\item One could follow \citet[][]{1993ApJ...413..447F} and expand the dark matter overdensity field $\delta_{\rm M}\equiv\rho_{\rm M}/\bar{\rho}_{\rm M}-1$ (with $\bar{\rho}_{\rm M}$ being the mean dark matter density) in a Taylor series:
\be
\rho_h=f^a_h\,\sum_{i}a_i\delta_{\rm M}^i{,}
\ee
with $f^a_h=\bar{N}_h/\langle \sum_{i}a_i\delta_{\rm M}^i \rangle$ and $a_i$ being the corresponding bias coefficients. The ensemble average over the quantity $X$: $\langle X\rangle$ can be approximated by the following average: $\sum_i X_i/N_{\rm cells}$ for sufficiently large  volumes (with $N_{\rm cells}$ being the number of cells dividing the entire volume).
This bias model has been proposed to augment the resolution of an $N$-body simulation, populating lower mass haloes than originally resolved in the simulation \citep[see][]{2013arXiv1310.3880A}.
\item Alternatively one could follow \citet[][]{Cen1993} and expand the series based on the logarithm of the density field (avoiding in this way negative densities allowed in the previous expansion):
\be
\rho_h=f^b_h\,\exp\left[\sum_{i}b_i\log\left(1+\delta_{\rm M}\right)^i\right]{,}
\ee
with $f^b_h=\bar{N}_h/\langle \exp\left[\sum_{i}b_i\log\left(1+\delta_{\rm M}\right)^i\right] \rangle$ and $b_i$ being the corresponding bias coefficients.
This model has been used for resolution augmentation of $N$-body simulations \citep[see][]{delaTorre2012}, further augmenting the catalogue with the halo occupation distribution (HOD) \citep[][]{2002ApJ...575..587B,2007ApJ...667..760Z,2013A&A...557A..54D}.
\item It has recently been found that the bias is very well fit by a compact relation including an exponential cut-off:  $\rho_h\propto\rho_{\rm M}^\alpha\,\exp\left[-\left(\frac{\rho_{\rm M}}{\rho_\epsilon}\right)^{\epsilon}\right]$ \citep[see][]{2014MNRAS.441..646N}.
This is a refined version of the thresholding step-function $\theta(\rho_{\rm M}-\rho_{\rm th})$ ($=0$ for $\rho_{\rm M}<\rho_{\rm th}$, $=1$ for $\rho_{\rm M}\geq\rho_{\rm th}$), which suppresses the generation of haloes below a density threshold $\rho_{\rm th}$ and was previously used in \citep[][]{patchy}. Such a model is in agreement with the \citet[][]{1974ApJ...187..425P} and peak-background split  picture \citep[][]{Kaiser84,bbks1986,Cole89,Mo96,Sheth01}. One can combine both descriptions in a single model:
\be
\rho_h=f_h\,\theta(\rho_{\rm M}-\rho_{\rm th})\,\rho_{\rm M}^\alpha\,\exp\left[-\left(\frac{\rho_{\rm M}}{\rho_\epsilon}\right)^{\epsilon}\right]{,}
\ee
with 
\be
\label{eq:numden}
f_h=\bar{N}_h/\langle\theta(\rho_{\rm M}-\rho_{\rm th})\,\rho_{\rm M}^\alpha\,\exp\left[-\left(\frac{\rho_{\rm M}}{\rho_\epsilon}\right)^{\epsilon}\right]\rangle{,}
\ee
 and \{$\rho_{\rm th},\alpha,\epsilon,\rho_\epsilon$\} the parameters of the model.
Nevertheless, the exponential cut-off ($\epsilon<0$) is not very sensible for the massive halo/galaxy population considered in this work. Hereby, the bias is mainly controled by $\alpha$ and $\rho_{\rm th}$. Different combinations of these parameters can lead to the same power spectrum (two-point statistics). Therefore, we need to introduce additional constraints, as we will discuss below. 
\end{enumerate}
 Additional bias is introduced by the approximate gravity solver and aliasing effects due to the gridding of the dark matter particles, when defining the dark matter density field $\rho_{\rm M}$.
We rely in this work on the (iii) bias expression and include additional bias effects in these parameters.

A non-local bias introduces a scatter in the local deterministic bias relations described above \citep[][]{Mcdonald09}. These effects have recently been found to be relevant \citep[][]{Saito14}, which is in line with the findings of \citet[][]{patchy}. In the latter work, the scatter is described within a stochastic bias relation through the negative binomial distribution function, as we do here. Therefore, stochastic bias may be regarded as an effective description of the non-local bias encoding our ignorance about it.  Nevertheless, such effects need to be addressed in more detail in  future works, as a stochastic description may lack some physical effects in the halo distribution.

\subsection{Halo probability distribution function}

The approach described so far to generate the halo distribution based on Eq.~\ref{eq:samp2} is not very efficient. One would need to scan the bias parameter space computing the correlation functions (say up to fourth order) for each set of values until the optimal set is found. A fast method has recently been developed to compute the bisepctrum in the squeezed-limit \citep[][]{Chiang2014}, which however, does not apply for this work. To circumvent these computational problems, we suggest to constrain the one-dimensional halo PDF.
In particular we will constrain the shape of the halo PDF, as its integral (the halo number density) will be imposed by $f_h$ (see Eq.~\ref{eq:numden}). 

Let us recap some concepts of statistical probability theory \cite[for a review see][]{2002PhR...367....1B}. 

\subsubsection{Statistical probability theory}

The higher-order moments of a field $\rho$ can be found by calculating the ensemble average of powers of that field $\rho$ over its probability distribution function ${\mathcal P}^1(\rho)$:
\begin{equation}
\mu_n \equiv \int \dd \rho\, {\mathcal P}^1(\rho) \,\rho^n=\langle\rho^n\rangle { , }
\end{equation} 
with $n$ being the order of the moment.

One can then define the moment generating function as the quantity:
\begin{equation}
  \label{eq:momgenfunc}
  \mathcal M(t)\equiv \sum_{n=0}^\infty \mu_n\frac{t^n}{n!} =\int \dd\rho \, {\mathcal P}^1(\rho) e^{t\rho}=\langle e^{t\rho}\rangle\,,
\end{equation} 
which yields the different moments by performing subsequent derivatives of it  at the origin $t=0$.

The cosmological $n$-point correlation functions are related to the cumulants or connected moments:
\begin{equation}
  \kappa_n \equiv\langle \rho^n\rangle_\cc {,}
\end{equation} 
defined by the cumulant generating function:
\begin{equation}
  \mathcal C(t)\equiv \sum_{n=1}^\infty \kappa_n \frac{t^n}{n!}{,}
\end{equation} 
and its relation to the moment generating function:
\begin{equation}
  \label{eq:momcum}
  \mathcal M(t)=\exp(\mathcal C(t)) {.}
\end{equation} 

By inverting Eq.~\ref{eq:momgenfunc}
using the inverse Laplace transform, one obtains the PDF as a function of the cumulants: 
\begin{equation}
{\mathcal P}^1(\rho)=\int^{\sqrt{-1}\infty}_{-\sqrt{-1}\infty}\frac{\dd t}{2\pi\sqrt{-1}}\exp\left(t\rho+\mathcal C(t)\right) {.}
\end{equation} 
This equation illustrates how the $n$-point correlation functions are encoded in the probability distribution function.

\subsubsection{Multivariate constraints}
\label{sec:assumps}

We have considered so far the univariate case.
 However, this alone does not ensure the correct higher-order statistics, as any distribution (for instance a Gaussian field) can be rank-ordered to fulfill a particular PDF \citep[see][]{weinberg,Sigad2000,Szapudi2004}.  
  The cosmic density field represents a multivariate statistical problem. In practice the statistical dimension  is given by the number of cells (voxels) conforming the three dimensional grid we use to model the whole cosmic volume under consideration \citep[for the more general multivariate expressions see][]{2002PhR...367....1B,kitaura_skewlog}.

One needs to constrain the PDF from the multivariate halo distribution. Following Eq.~\ref{eq:samp2} we need to assume that at scales larger than few Mpc (i.e., larger than the diameter of the largest haloes):
\begin{itemize}
\item The approximate gravity solver (low $N$-body resolution or perturbation theory based method) accurately models the higher-order statistics of the dark matter density field.
\item The bias model accurately connects the dark matter phase-space distribution with the halo distribution.
\end{itemize}
In this way, we scan the parameter space spanned by the bias parameters, in our case: $\{\rho_{\rm th},\alpha,\epsilon,\rho_\epsilon,\beta\}$, additionaly fixing the number density of the halo population in which we are interested: $\bar{N}_h=\langle \rho_h\rangle\leftarrow\xi^h_1$, to match the power spectrum: $P_h(k)\leftarrow\xi^h_2$
and the halo PDF:
${\mathcal P}^1_h\left(B\left(\rho_h|\rho_{\rm M}\right)\right)\leftarrow \{\xi^h_1,\xi^h_2,\xi^h_3,\xi^h_4,\dots\}$ obtained from $N$-body simulations. The computation of the halo PDF is trivial and fast in contrast to the calculation of higher-order correlation functions. 
This can be summarised with the following sampling process:
\begin{equation}
N_h\curvearrowleft {\mathcal P}(N_h|B(\rho_h|\rho_{\rm M},\bar{N}_h,P_h(k),{\mathcal P}^1_h)) {.}
\end{equation} 
The accuracy of our method depends on the level of precision in which each of the two above mentioned  conditions  are fulfilled.
Recently, rank-ordering has been applied to dark matter density fields within second order LPT (2LPT) demonstrating that this approximation encaptures the right matter statistics on scales larger than 8 $h^{-1}$ Mpc, based on calculations of the bispectrum  \citep[see][]{Leclercq2013}. LPT was also shown to accurately model the three-point correlation function on scales relevant to BAOs in configuration space \citep[][]{2014MNRAS.437.2594W}.   The limitations of LPT can be improved with the spherical collapse model \citep[see][and the description of the \textsc{patchy}-code below]{Neyrinck2013,KitauraHess2013}. 
We stress that the advantage of the above probabilistic formulation goes beyond of producing mock catalogues, as it permits us also for statistical inference analysis \citep[see e.~g.~][]{kitaura,kitaura_log,jasche_hamil,kitaura_lyman,kitaura_kigen}.

\section{Numerical experiments}
\label{sec:results}

We present in this section numerical experiments to validate our method. 

\subsection{Reference halo sample: the BigMutiDark simulation}
\label{sec:BMD}

To calibrate our method, we use a reference halo catalogue at redshift $z=0.577$ extracted  from one of the  {BigMultiDark} (\textsc{BigMD}) simulations \citep[He{\ss} et al., in prep.; ][]{Klypin2013}, which was performed using \textsc{gadget-2} \cite[][]{gadget2} with $3840^3$ particles on a volume of $(2500\, h^{-1}$ Mpc)$^3$ assuming $\Lambda$CDM-cosmology with \{$\Omega_{\rm M}=0.29,\Omega_{\rm K}=0,\Omega_\Lambda=0.71,\Omega_{\rm B}=0.047,\sigma_8=0.82,w=-1,n_s=0.95$\} and a Hubble constant ($H_0=100\,h$ km s$^{-1}$ Mpc$^{-1}$) given by  $h=0.7$. Haloes were defined based on density peaks including substructures  using the Bound Density Maximum (BDM) halo finder \citep[][]{klypin1997} and then selected according to a maximum circular velocity larger than 350 km s$^{-1}$ to match the number density of BOSS CMASS galaxies \citep[][]{nuza2013,Klypin2013}. For the impact of these selection criteria in the clustering and scale-dependent bias (Prada et al., in prep.).

\subsection{The \textsc{patchy}-code}

 To maximise the efficiency of the method we rely on augmented Lagrangian perturbation theory (ALPT) splitting the description of the large-scale structure into a long range and a short range component modelled by 2LPT and the spherical collapse model, respectively  \citep[see][]{KitauraHess2013}. We note, however, that the method described above can be applied to increase the resolution of $N$-body simulations, and thus, it can be regarded as an improved method with an extended parametrised bias model with respect to other approaches like the ones presented in \citet[][]{delaTorre2012,2013arXiv1310.3880A}. It has been shown with the \textsc{patchy}-code that the non-linear bias model also adopted in this work can compensate for the missing power of the perturbative approach and redshift space distortions can be accurately modelled with ALPT and a dispersion term \citep[see][]{patchy}.
We have extended  \textsc{patchy} to be out-of-core and work with an arbitrary number of chunks to reduce the memory requirements  below 28 Gb RAM for 1024$^3$   particles (or cells), allowing to simulate the distribution of LRG-like galaxies in volumes of (2.5 $h^{-1}$\,Gpc)$^3$. The new version of \textsc{patchy} randomly assigns the positions of dark matter particles to haloes, whenever there are more dark matter particles than haloes in a given cell. Otherwise the position within the cell is uniform randomly assigned. This case does, however, occur only for a small fraction of cells  considering the particular halo population of this work. 

\begin{figure}
\begin{tabular}{c}
\includegraphics[width=7.cm]{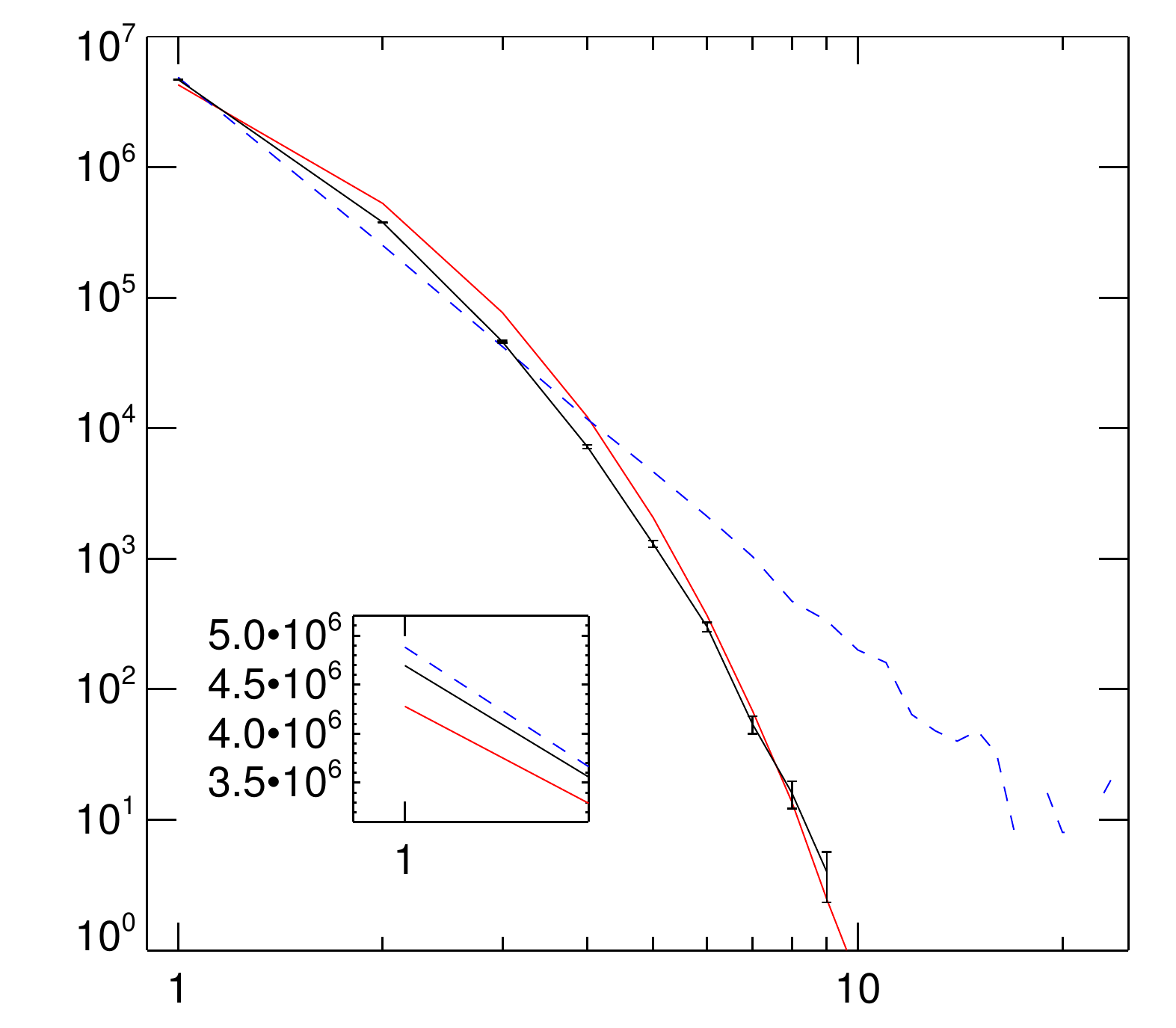}
\put(-200,90){\rotatebox[]{90}{$N_{\rm cells}$}}
\put(-110,-5){$N_{\rm haloes}$}
\end{tabular}
\caption{\label{fig:PDF} Halo PDF showing the number of cells $N_{\rm cells}$ hosting the same number of haloes $N_{\rm haloes}$ with a cell resolution of $(2.5 \,h^{-1}\, {\rm Mpc})^3$. In red \textsc{patchy} (with $\rho_{\rm th}^{\rm High}$) and in black \textsc{BigMD} including the error-bars extracted from  the \textsc{patchy} mocks. In addition one \textsc{patchy} mock in dashed-blue (with $\rho_{\rm th}^{\rm Low}$) ignoring the shape of the halo PDF.  The latter has been multiplied by eight to compensate for the eight times smaller volume, and hence, about eight times smaller number of haloes. The number density (which determines the integral of the PDF) is the same in all catalogues (see Eq.~\ref{eq:numden}).  The insert shows the first bin of the halo PDF for cells containing only one halo.   }
\end{figure}

\begin{figure*}
\begin{tabular}{cc}
\hspace{-.25cm}
\includegraphics[width=7.5cm]{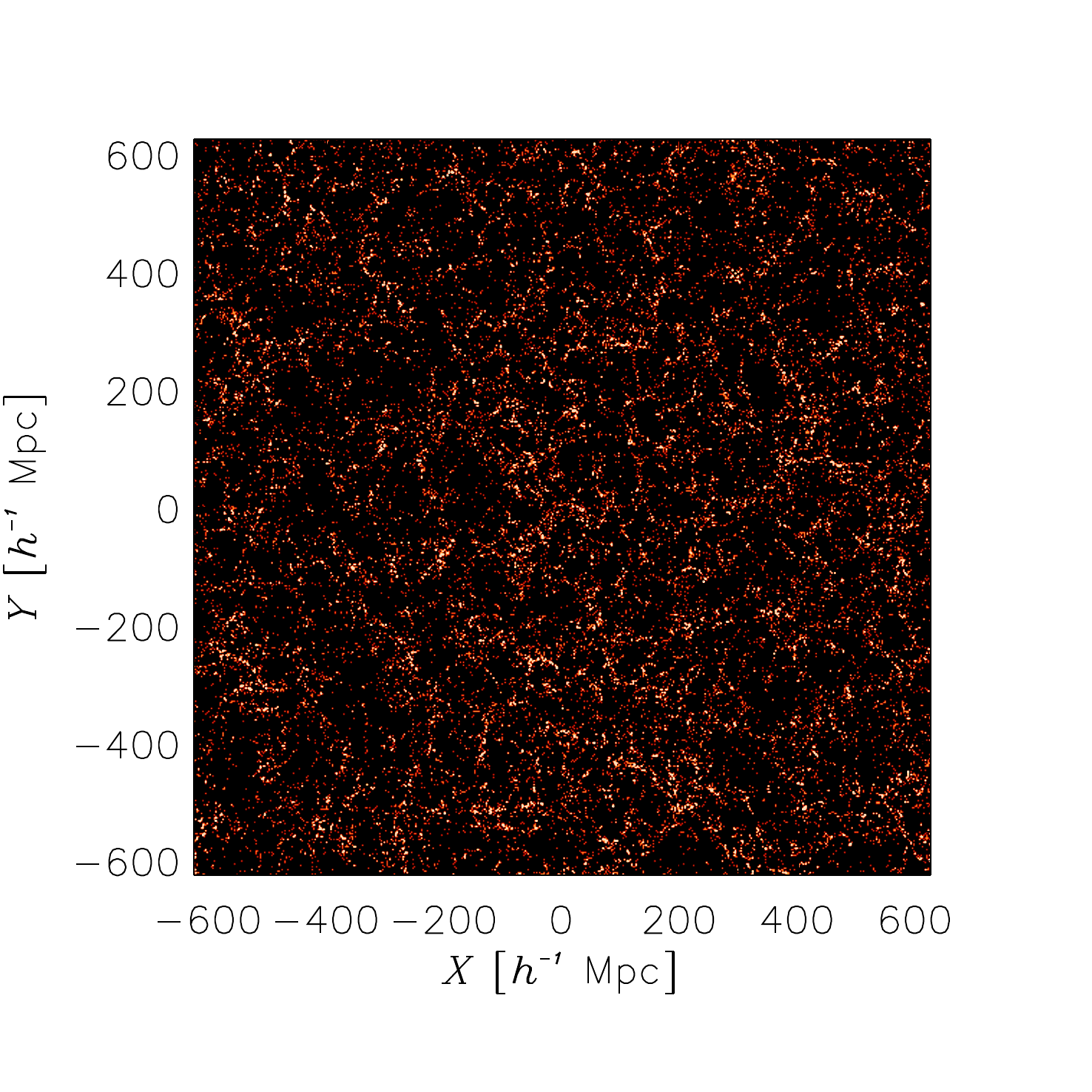}

\hspace{-2.cm}
\includegraphics[width=7.5cm]{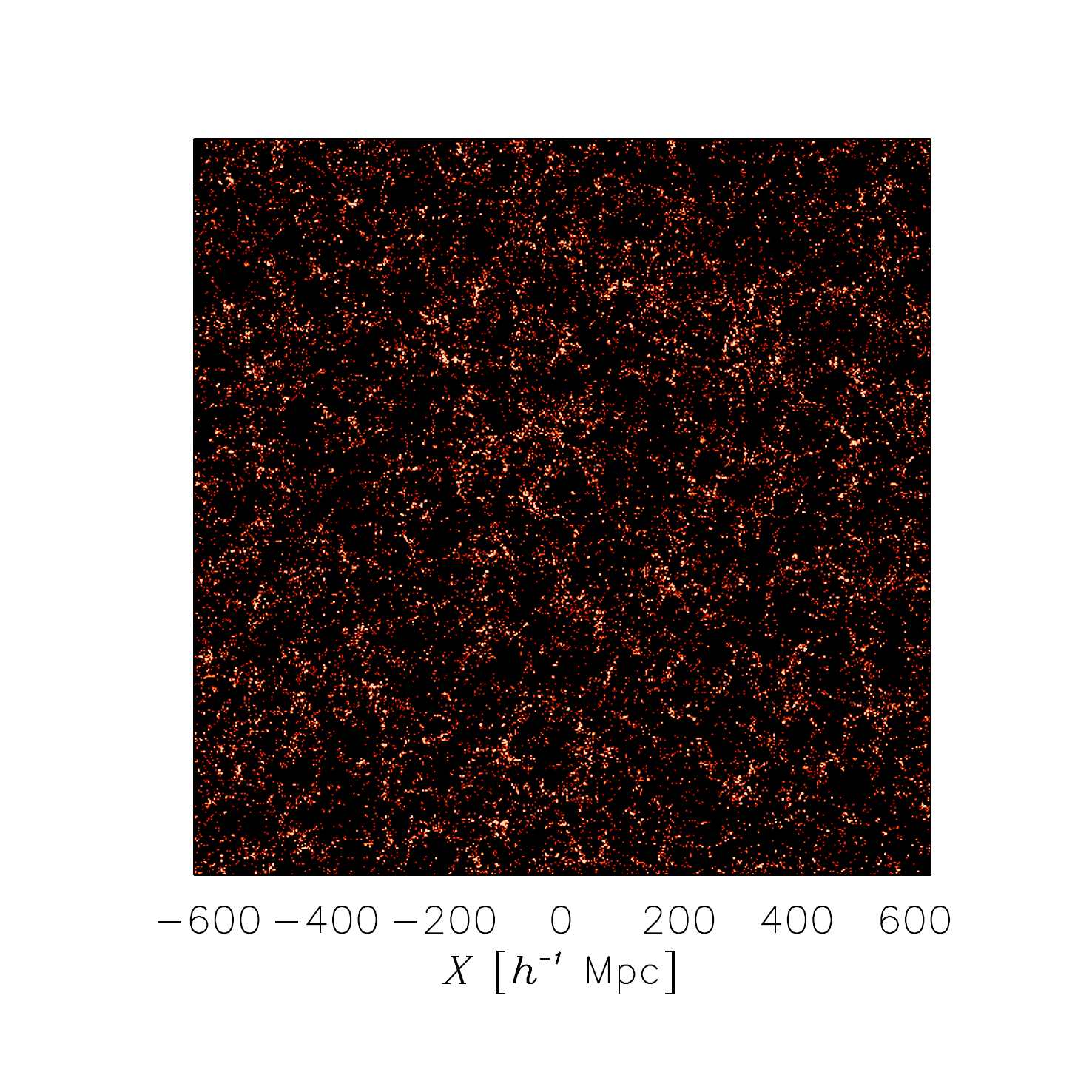}
\hspace{-2.cm}
\includegraphics[width=7.5cm]{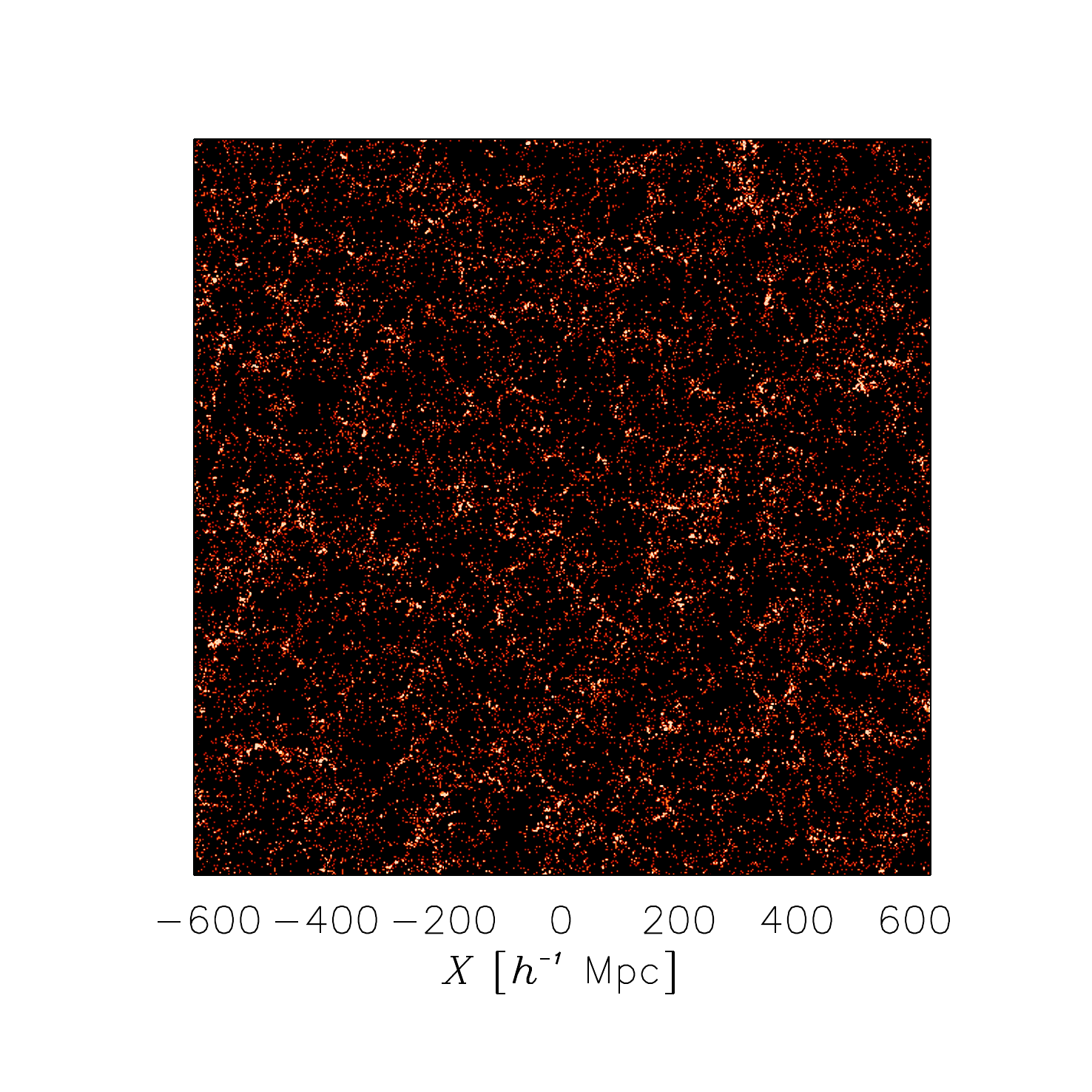}
\vspace{-.8cm}
\end{tabular}
\caption{\label{fig:hal}
Halo overdensity slices of thickness 20 $h^{-1}$ Mpc taking a subvolume of  (1250 $h^{-1}$ Mpc)$^3$ from the \textsc{BigMD} $N$-body simulation on the left, from one \textsc{patchy} mock including a fit of the halo PDF (with $\rho_{\rm th}^{\rm High}$) in the middle, and from one \textsc{patchy} mock ignoring the shape of the halo PDF (with $\rho_{\rm th}^{\rm Low}$) on the right. Lighter regions indicate a larger halo number count.}
\end{figure*}

\subsection{Bias parameters and statistical constraints}

The parameters of our model $\{\rho_{\rm th},\alpha,\epsilon,\rho_\epsilon,\beta\}$ are constrained to fit the power spectrum better than 2\% in the range $k=[0.07,0.35]\,h\,{\rm Mpc}^{-1}$ (cosmic variance dominates on larger scales: $k\lsim0.07\,h\,{\rm Mpc}^{-1}$)  and to accurately reproduce the tails of the halo PDF, i.e., matching the low-end better than 10\%  and being statistically compatible within 1-sigma at the high-end (around the maximum number of haloes per cell). We find that in this way we obtain also a reasonable fit of the halo PDF in the whole range, with deviations being in general within 50\%. Our bispectrum statistics analysis has shown that these deviations are not severe (see below). To achieve better fits we would need to match the low-end of the halo PDF to higher accuracy, as small deviations in the number of cells containing one halo can lead to larger deviations in cells containing higher number of haloes (see insert in Fig.~\ref{fig:PDF}). There should also be limitations due to the assumptions adopted in \S \ref{sec:assumps}. This should be further investigated in future works.
Nevertheless, a massive parameter estimation approach to find the optimal parameters to match the BOSS-CMASS galaxy population will be presented in a forthcoming paper (Scoccola et al., in prep.).
We note that the right number density is imposed through Eq.~\ref{eq:numden}.

A multiscale approach has been adopted in this work to efficiently scan the bias parameter space. 
We start with small volumes of $(312.5 \,h^{-1}\,{\rm Mpc})^3$ on grids with $128^3$ cells and refine the parameters increasing the volumes by factors of eight with constant resolution, until we reach the final volume of (2500 $h^{-1}\, {\rm Mpc}$)$^3$ with $1024^3$ cells.  
The majority of the tests were performed on volumes of (1250 $h^{-1}\, {\rm Mpc}$)$^3$ and grids of $512^3$, since the statistical uncertainties are already low at these volumes and they can be computed much faster than the full volume of the \textsc{BigMD} simulation.
We show in this study one of such mocks with the following parameters: $\alpha=1.68$ and $\delta_{th}=\rho_{\rm th}/\bar{\rho}-1=-0.3$, which ignores the shape of the halo PDF from the \textsc{BigMD} simulation and solely fits the halo number density (integral of the halo PDF) and the power spectra in real and redshift space (see \S~\ref{sec:pkandbk}). This setting permits the existence of haloes in low density regions,  we therefore will refer to this sample as the \textsc{patchy} mock with $\rho_{\rm th}^{\rm Low}$. The calculations based on this mock are represented by dashed-blue curves throughout the paper.
In addition, we show results from a set of 20 \textsc{patchy} mockswith parameters chosen to jointly fit the power spectrum and the halo PDF ($\rho_{\rm th}^{\rm High}$), using the same cosmology, number density and volume as the \textsc{BigMD} simulation described in \S~\ref{sec:BMD}. A $\sim$53 times lower number of dark matter particles (1024$^3$) was used for the \textsc{patchy} mocks.  We find that a higher thresholding is crucial to fit the PDF  (see Fig.~\ref{fig:PDF}). The following parameters are found: $\alpha=0.365$ and $\delta_{th}=\rho_{\rm th}/\bar{\rho}-1=1.82$, and therefore we will refer to this sample as the \textsc{patchy} mocks with $\rho_{\rm th}^{\rm High}$.
The calculations based on these mocks are represented by red curves throughout the paper.
We find for the parameter modelling the deviation from Poissonity $\beta=0.35$. In this way we manage to enhance the power towards small scales (high $k$) fitting the power spectrum of the \textsc{BigMD} simulation (see Fig.~\ref{fig:pk}). The exponential cut-off parameters are chosen as: $\epsilon=-2$, $\rho_{\epsilon}=0.3$. We note, however, that these parameters have a limited impact on the statistics given the particular halo population we are looking at. We expect that they will become more relevant when looking at lower halo masses.
 
A visual comparison between the halo field from the \textsc{BigMD} simulation and from the \textsc{patchy} mocks is shown in Fig.~\ref{fig:hal}.  There are no obvious deviations between the \textsc{BigMD} simulation and the \textsc{patchy} mocks  other than a different spatial distribution of haloes due to the different initial random seed perturbations used for each case. A more careful visual inspection reveals that the \textsc{patchy} mock with $\rho_{\rm th}^{\rm Low}$ exhibits a larger number of cells with low halo number counts filling the voids with respect to the \textsc{patchy} mock with $\rho_{\rm th}^{\rm High}$ and the   \textsc{BigMD} simulation. This is expected as the density threshold is set to a lower value. We cannot distinguished by eye the larger saturation of the peaks. This is however shown in the PDF, where one can see that the \textsc{patchy} mock with $\rho_{\rm th}^{\rm Low}$ reaches a factor $\sim3$ times larger number of haloes per cell than the  \textsc{patchy} mock with $\rho_{\rm th}^{\rm High}$ or the   \textsc{BigMD} simulation  (see Fig.~\ref{fig:PDF}).   We proceed with a more quantitative statistical analysis.

\subsection{Power spectrum and bispectrum}
\label{sec:pkandbk}

In this section we aim to compare the halo power spectrum and halo bispectrum predictions for \textsc{patchy} mocks and  \textsc{BigMD} $N$-body simulation in real and redshift space.

The halo power spectrum $P$ and the halo bispectrum $B$ are the two- and three-point correlation functions in Fourier space. Given the Fourier transform of the halo overdensity, $\delta_h({\bf x})\equiv\rho_{h}({\bf x})/{\bar \rho}_h-1$,
\begin{equation}
\delta_h({\bf k})={\int} d^3{\bf x}\,\delta_h({\bf x})\exp(-i{\bf k}\cdot{\bf x}),
\end{equation}
where $\rho_h({\bf x})$ is the number density of objects and $\bar{\rho}_h$ its mean value, the halo power spectrum and halo bispectrum are defined as,
\begin{eqnarray}
\langle \delta_h({\bf k}) \delta_h({{\bf k}'})\rangle&\equiv&(2\pi)^3P( k)\delta^D({\bf k}+{\bf k}'),\\
\langle \delta_h({{\bf k}_1}) \delta_h({{\bf k}_2}) \delta_h({{\bf k}_3})\rangle&\equiv&(2\pi)^3B({\bf k}_1, {\bf k}_2)\delta^D({\bf k}_1+{\bf k}_2+{\bf k}_3)\,{,}\nonumber\\
\end{eqnarray}
with $\delta^D$ being the Dirac delta function. Note that the bispectrum is only well defined when the set of $k$-vectors, $k_1$, $k_2$ and $k_3$ close to form a triangle, ${\bf k}_1+{\bf k}_2+{\bf k}_3={\bf 0}$.
It is common to define the reduced bispectrum $Q$ as, 
\begin{equation}
Q(\alpha_{12}|{\bf k}_1,{\bf k}_2)\equiv\frac{B({\bf k}_1,{\bf k}_2)}{P(k_1)P(k_2)+P(k_2)P(k_3)+P(k_1)P(k_3)}.
\end{equation}
where $\alpha_{12}$ is the angle between ${\bf k}_1$ and ${\bf k}_2$. Although this quantity does not have any additional information to the power spectrum and bispectrum, it has been historically used as a measurement of the hierarchical amplitude of the bispectrum. This quantity is independent of the overall scale $k$ and redshift at large scales and for a power spectrum that follows a power law. Moreover, it presents a characteristic ``U-shape" predicted by the gravitational instability. 
Mode coupling and power law deviations in the actual power spectrum induce a slight scale- and time-dependency in this quantity. However,  in practice it has been observed that at scales of order $k\sim0.1\,h\,{\rm Mpc}^{-1}$ the reduced bispectrum does not present a high variation in its amplitude.

\begin{figure*}
\begin{tabular}{cc}
\hspace{-0.25cm}
\includegraphics[width=9.cm]{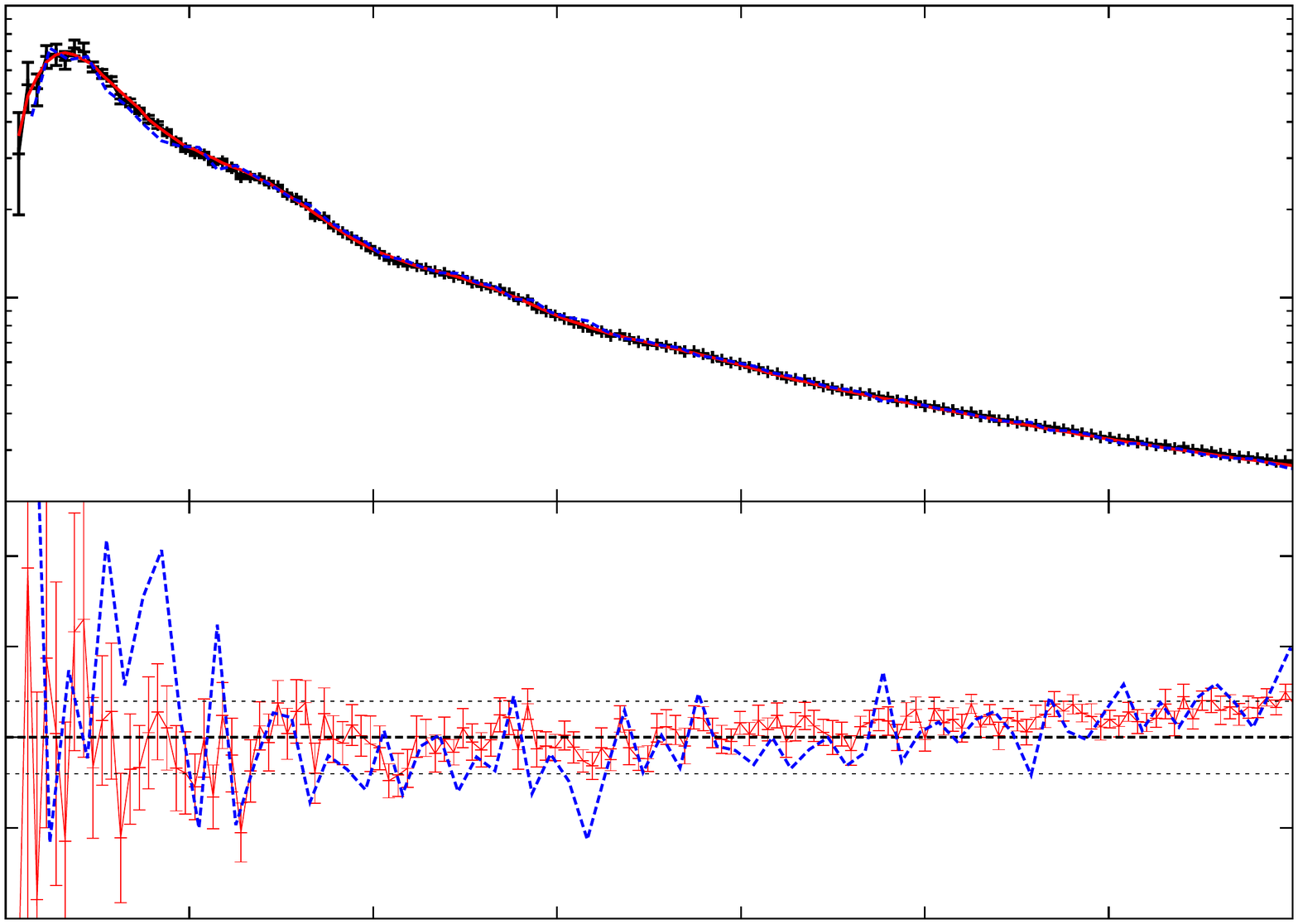}

\put(-80,160){\rotatebox[]{0}{\text{ real space}}}

\put(-252,140){\rotatebox[]{90}{\text{\tiny $P(k)\times10^{4} \,[(h^{-1}\,{\rm Mpc})^3]$}}}
\put(-262,70){\rotatebox[]{90}{\text{\tiny $P_{\rm BigMD}(k)/$}}}
\put(-252,70){\rotatebox[]{90}{\text{\tiny $P_{\rm Patchy}(k)$}}}

\put(-230,25){\text{\tiny0}}
\put(-206,25){\text{\tiny0.05}}
\put(-173,25){\text{\tiny0.10}}
\put(-143,25){\text{\tiny0.15}}
\put(-113,25){\text{\tiny0.20}}
\put(-83,25){\text{\tiny0.25}}
\put(-52,25){\text{\tiny0.30}}
\put(-23,25){\text{\tiny0.35}}

\put(-239,180){\text{\tiny\tiny10}}
\put(-236,132){\text{\tiny\tiny1}}

\put(-239.5,30){\text{\tiny\tiny0.9}}
\put(-241.7,45){\text{\tiny\tiny0.95}}
\put(-236,60){\text{\tiny\tiny1}}
\put(-241.7,75){\text{\tiny\tiny1.05}}
\put(-239.5,90){\text{\tiny\tiny1.1}}

\put(-140,15){\text{\tiny $k \,[h\,{\rm Mpc}^{-1}]$}}
\hspace{-0.25cm}
\includegraphics[width=9.cm]{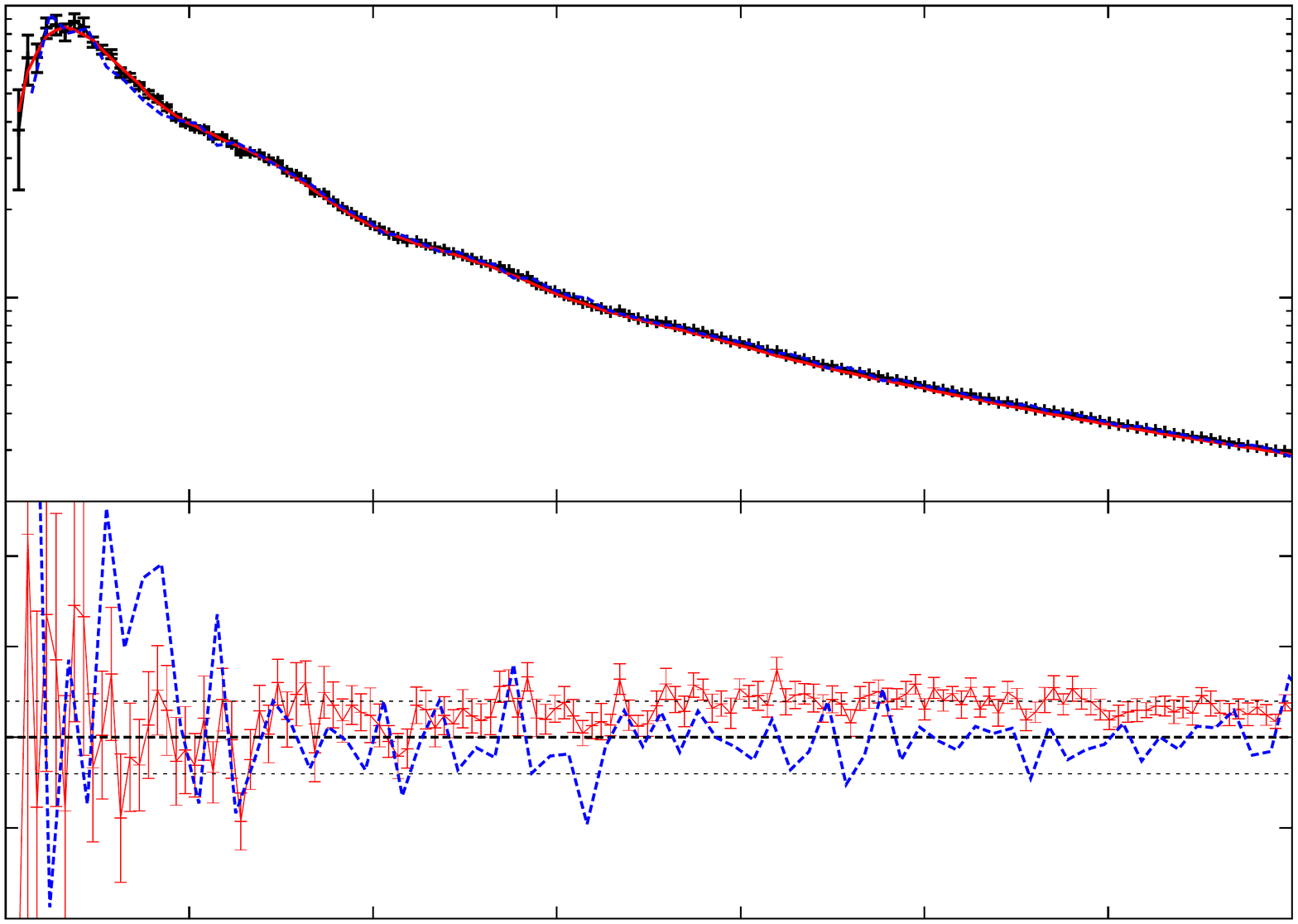}

\put(-80,160){\rotatebox[]{0}{\text{ redshift space}}}

\put(-252,140){\rotatebox[]{90}{\text{\tiny $P^{(0)}(k)\times10^{4} \,[(h^{-1}\,{\rm Mpc})^3]$}}}
\put(-264,70){\rotatebox[]{90}{\text{\tiny $P^{(0)}_{\rm BigMD}(k)/$}}}
\put(-253,70){\rotatebox[]{90}{\text{\tiny $P^{(0)}_{\rm Patchy}(k)$}}}

\put(-239,180){\text{\tiny\tiny10}}
\put(-236,132){\text{\tiny\tiny1}}

\put(-239.5,30){\text{\tiny\tiny0.9}}
\put(-241.7,45){\text{\tiny\tiny0.95}}
\put(-236,60){\text{\tiny\tiny1}}
\put(-241.7,75){\text{\tiny\tiny1.05}}
\put(-239.5,90){\text{\tiny\tiny1.1}}

\put(-230,25){\text{\tiny0}}
\put(-206,25){\text{\tiny0.05}}
\put(-173,25){\text{\tiny0.10}}
\put(-143,25){\text{\tiny0.15}}
\put(-113,25){\text{\tiny0.20}}
\put(-83,25){\text{\tiny0.25}}
\put(-52,25){\text{\tiny0.30}}
\put(-23,25){\text{\tiny0.35}}

\put(-140,15){\text{\tiny $k \,[h\,{\rm Mpc}^{-1}]$}}
\vspace{-.5cm}
\end{tabular}
\caption{\label{fig:pk}
Halo power spectrum for  \textsc{BigMD} $N$-body simulation (black symbols in the top sub-panels) and for \textsc{patchy} mocks (red lines with $\rho_{\rm th}^{\rm High}$ and dashed-blue lines  with $\rho_{\rm th}^{\rm Low}$). Left panel shows real space power spectrum and right panel the redshift space monopole power spectrum. The power spectrum for the \textsc{patchy} mocks with $\rho_{\rm th}^{\rm High}$ corresponds to the average over 20 independent realizations. Only the errors corresponding to the \textsc{BigMD} $N$-body simulations are shown for clarity. In the  bottom sub-panels the relative deviation between \textsc{BigMD} and \textsc{patchy} is shown. Black dotted lines indicate $2\%$ deviation.}
\end{figure*}

In order to measure the Fourier statistics from a set of haloes in a box with periodic boundary conditions, we start by discretising the box in grid cells. We use $512^3$ grid cells, with a grid-cell resolution of  $4.88\,h^{-1}\,{\rm Mpc}$, and we assign haloes to the grid cells according to the Cloud-in-Cell (CiC) prescription. We apply the Fourier transform of the density field using fftw\footnote{Fastest Fourier Transform of the West: www.fftw.org} and we correct the resulting field by deconvolving the effect of the grid assignment \citep[][]{Jing2005}. Given that the size of the box is $L_b=2.5\,h^{-1}\,{\rm Gpc}$, the $k$ fundamental is $k_f=2\pi/L_b=2.51\times10^{-3}\,h\,{\rm Mpc}^{-1}$.

In order to obtain the power spectrum we radially bin the $k$-modes, with a bin size of $k_f$, and we average over all possible $k$-directions.

We use the real part of $\langle \delta_{{\bf k}_1} \delta_{{\bf k}_2} \delta_{{\bf k}_3}\rangle$ as our bispectrum measurement for those set of ${\bf k}_i$-vectors that form a  triangle. Since there are a huge number of possible triangular configurations, in this paper we focus on analysing those with $k_1/k_2=1$ and $k_1/k_2=2$, for a particular values of $k_1$. We present the bispectrum measurement as a function of $k_3$, which  sweeps from $|{\bf k}_1-{\bf k}_2|$ to $|{\bf k}_1+{\bf k}_2|$. Our choice for binning the bispectrum is similar to the power spectrum. We bin $k_1$ and $k_3$ in $k_f$ bins: $\Delta k_1=\Delta k_3=k_f$. In addition, $k_2$ is binned also in $k_f$ bins for those triangles with $k_1/k_2=1$. On the other hand, $k_2$ is binned in 2 times $k_f$ when $k_2/k_1=2$ in order to cover all the available $k$-space. Generically, we can write $\Delta k_2=(k_2/k_1)\Delta k_1$.

\begin{figure*}
\begin{tabular}{cccc}
\includegraphics[width=4.5cm]{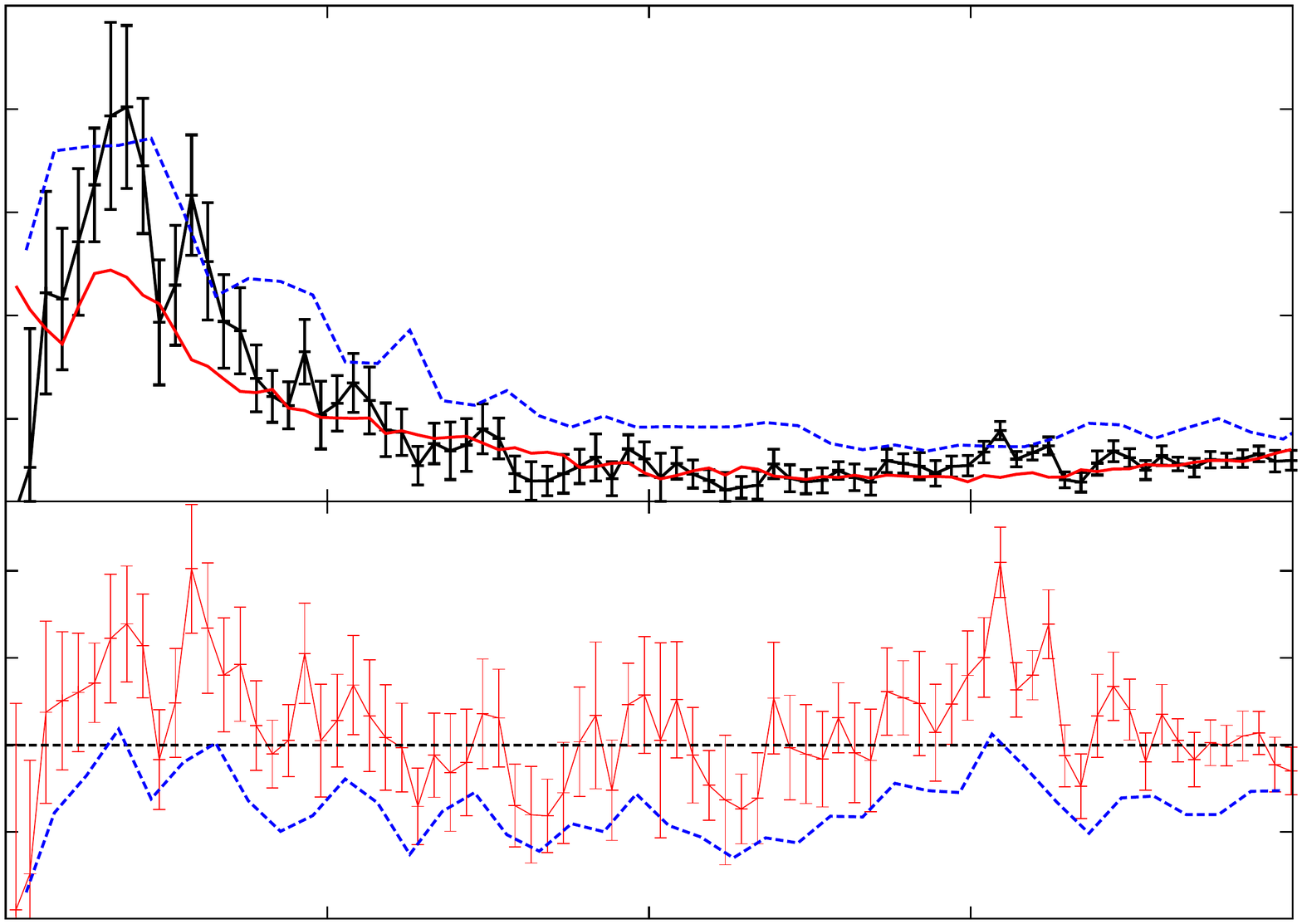}
\put(-85,80){\text{\tiny $k_2=k_1=0.1\, h\,{\rm Mpc}^{-1}$}}
\put(-145,70){\rotatebox[]{90}{\text{\tiny\tiny $B(k)\times10^{9}$}}}
\put(-135,70){\rotatebox[]{90}{\text{\tiny\tiny $[(h^{-1}\,{\rm Mpc})^{6}]$}}}
\put(-145,27){\rotatebox[]{90}{\text{\tiny\tiny $B_{\rm BigMD}(k)/$}}}
\put(-135,27){\rotatebox[]{90}{\text{\tiny\tiny $B_{\rm Patchy}(k)$}}}
\put(-122.7,55.5){\text{\tiny\tiny0.5}}
\put(-119,63.5){\text{\tiny\tiny1}}
\put(-122.7,72){\text{\tiny\tiny1.5}}
\put(-119,80){\text{\tiny\tiny2}}
\put(-122.7,89){\text{\tiny\tiny2.5}}

\put(-119,15.){\text{\tiny\tiny0}}
\put(-122.7,21.5){\text{\tiny\tiny0.5}}
\put(-119,28.3){\text{\tiny\tiny1}}
\put(-122.7,35.7){\text{\tiny\tiny1.5}}
\put(-119,43){\text{\tiny\tiny2}}

\hspace{-0.2cm}

\includegraphics[width=4.5cm]{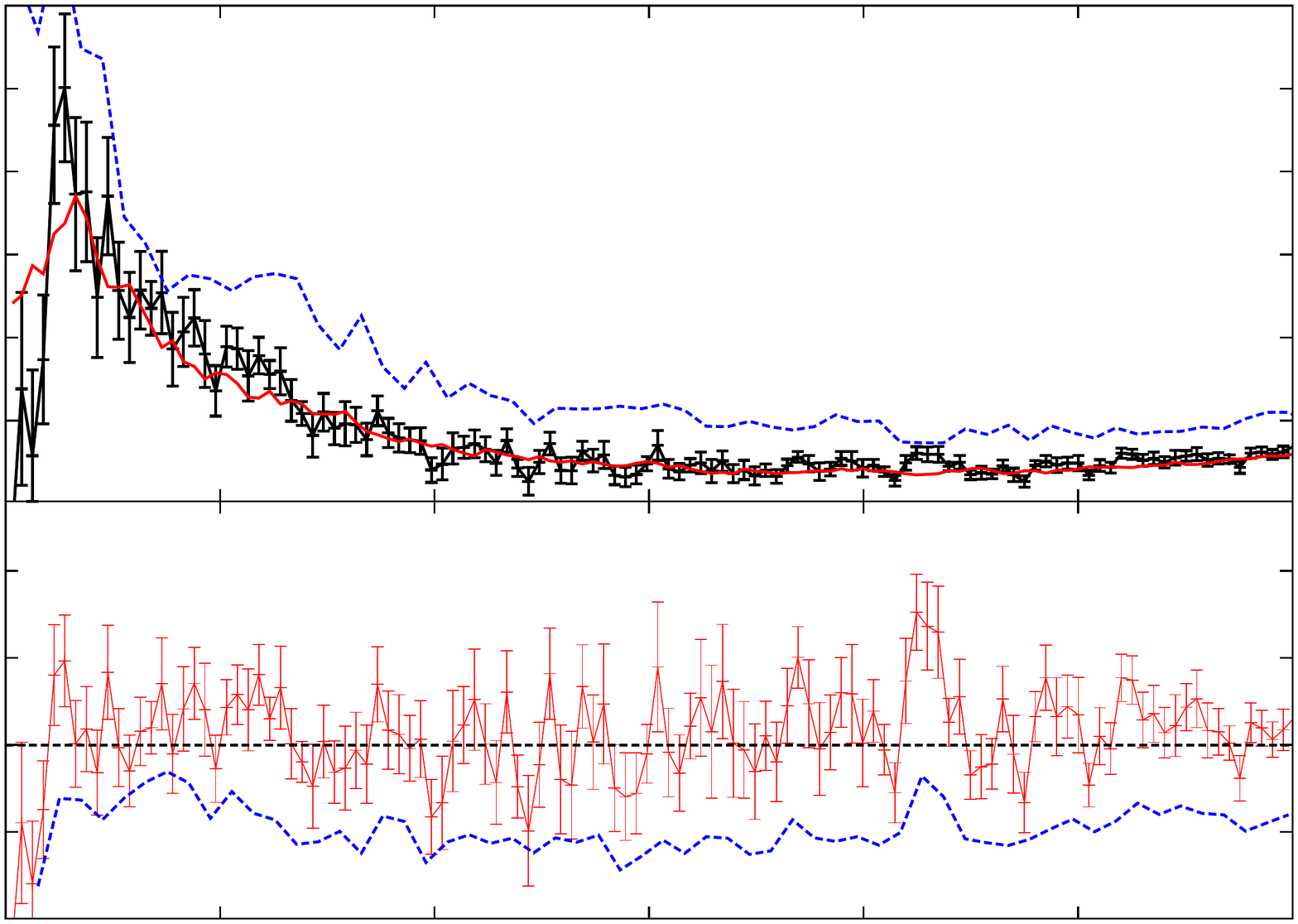}
\put(-88,80){\text{\tiny $k_2=k_1=0.15\, h\,{\rm Mpc}^{-1}$}}

\put(-122.7,55.){\text{\tiny\tiny0.2}}
\put(-122.7,62.){\text{\tiny\tiny0.4}}
\put(-122.7,68.5){\text{\tiny\tiny0.6}}
\put(-122.7,75.5){\text{\tiny\tiny0.8}}
\put(-122.7,82.5){\text{\tiny\tiny1.0}}
\put(-122.7,89){\text{\tiny\tiny1.2}}

\put(-119,15.){\text{\tiny\tiny0}}
\put(-122.7,21.5){\text{\tiny\tiny0.5}}
\put(-119,28.3){\text{\tiny\tiny1}}
\put(-122.7,35.7){\text{\tiny\tiny1.5}}
\put(-119,43){\text{\tiny\tiny2}}

\hspace{-0.2cm}
\includegraphics[width=4.5cm]{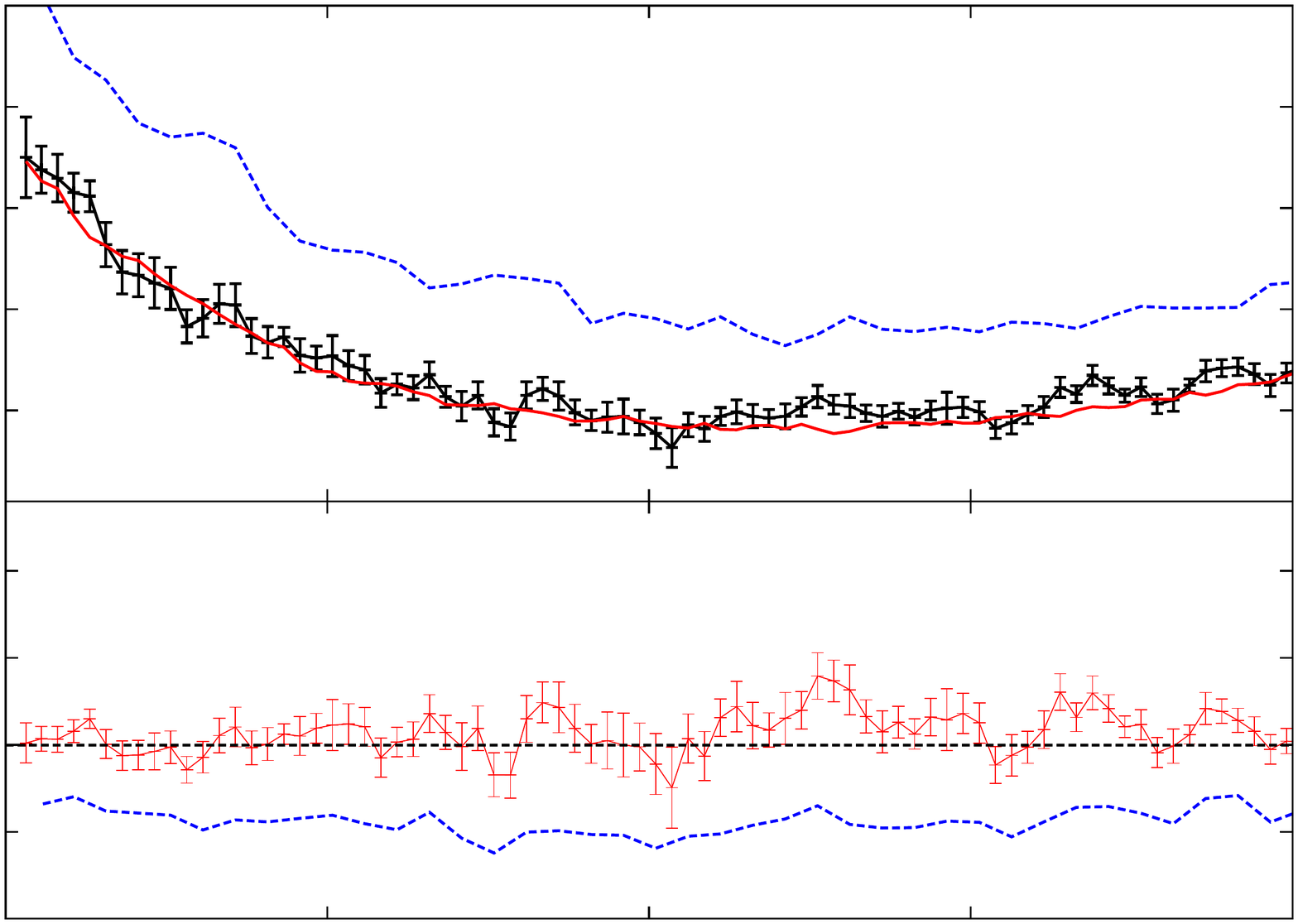}

\put(-122.7,56.5){\text{\tiny\tiny0.1}}
\put(-122.7,64.5){\text{\tiny\tiny0.2}}
\put(-122.7,72.5){\text{\tiny\tiny0.3}}
\put(-122.7,81){\text{\tiny\tiny0.4}}
\put(-122.7,89){\text{\tiny\tiny0.5}}

\put(-88,80){\text{\tiny $k_2=2k_1=0.2\, h\,{\rm Mpc}^{-1}$}}

\put(-119,15.){\text{\tiny\tiny0}}
\put(-122.7,21.5){\text{\tiny\tiny0.5}}
\put(-119,28.3){\text{\tiny\tiny1}}
\put(-122.7,35.7){\text{\tiny\tiny1.5}}
\put(-119,43){\text{\tiny\tiny2}}

\hspace{-0.2cm}
\includegraphics[width=4.5cm]{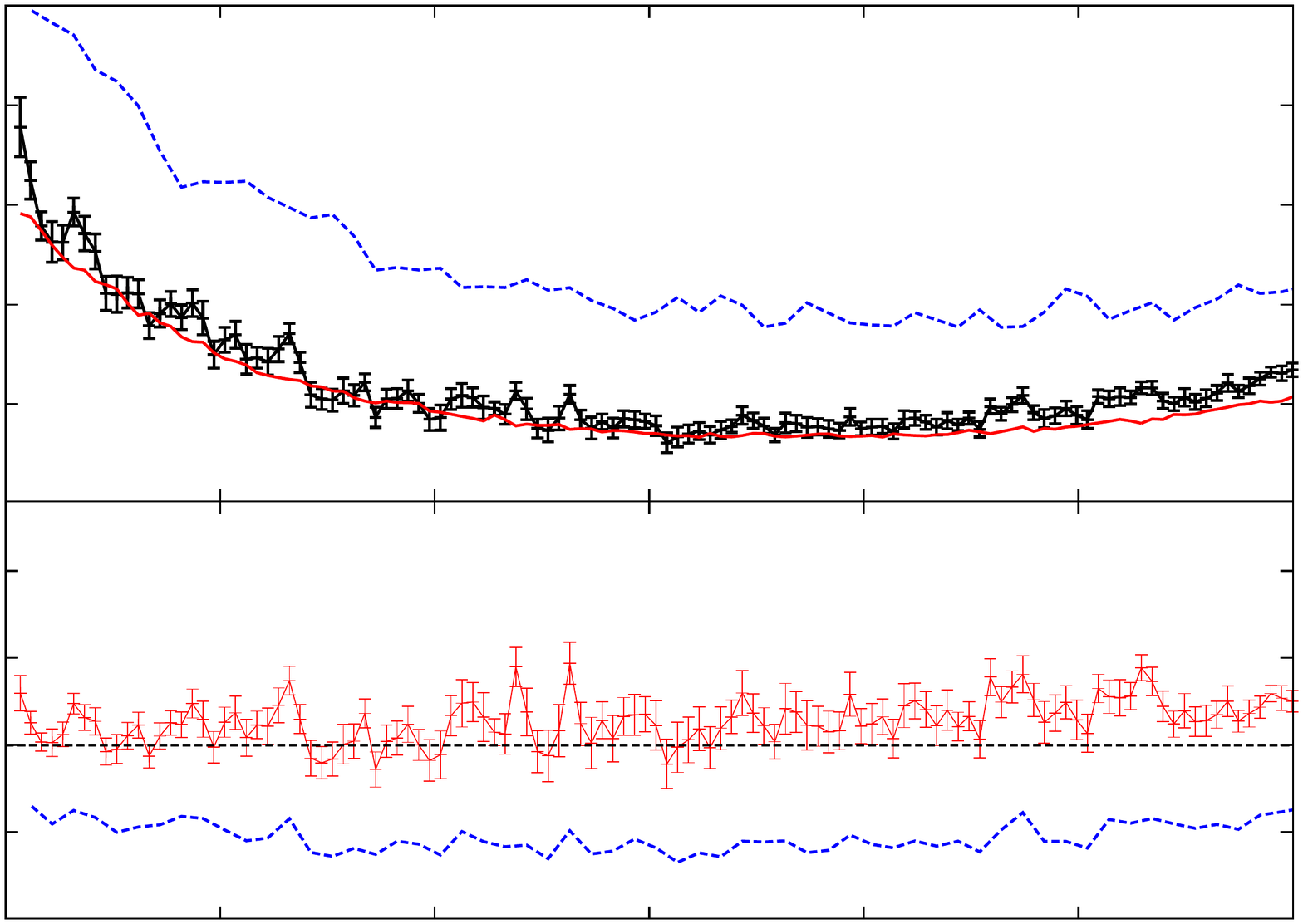}

\put(-122.7,56.5){\text{\tiny\tiny0.4}}
\put(-122.7,65.){\text{\tiny\tiny0.8}}
\put(-122.7,73){\text{\tiny\tiny1.2}}
\put(-122.7,81.6){\text{\tiny\tiny1.6}}
\put(-119,89){\text{\tiny\tiny2}}

\put(-91,80){\text{\tiny $k_2=2k_1=0.3\, h\,{\rm Mpc}^{-1}$}}

\put(-119,15.){\text{\tiny\tiny0}}
\put(-122.7,21.5){\text{\tiny\tiny0.5}}
\put(-119,28.3){\text{\tiny\tiny1}}
\put(-122.7,35.7){\text{\tiny\tiny1.5}}
\put(-119,43){\text{\tiny\tiny2}}

\vspace{-.5cm}
\\
\hspace{-0.1cm}

\includegraphics[width=4.5cm]{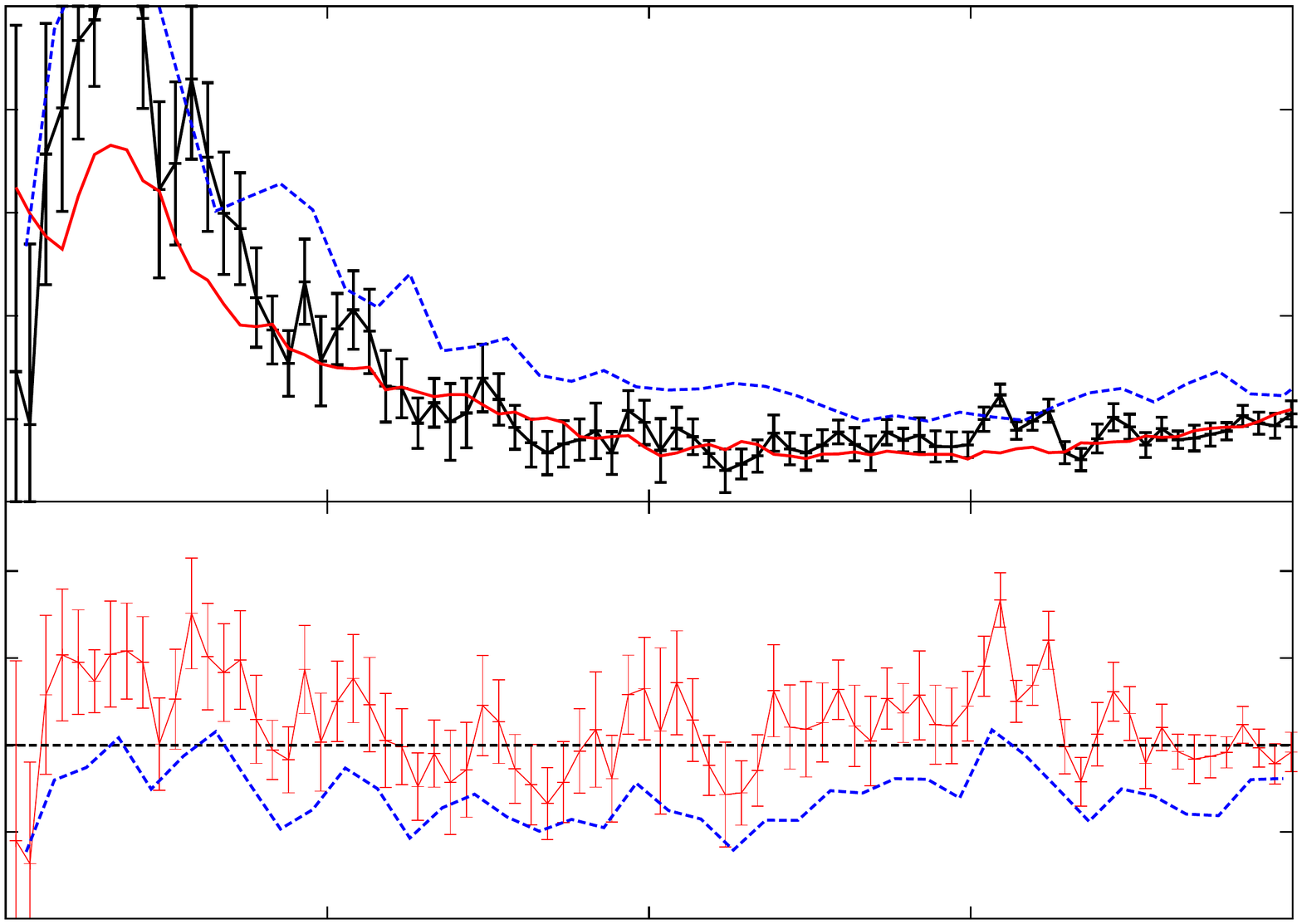}
\put(-80,2){\text{\tiny\tiny $k_3 \,[h\,{\rm Mpc}^{-1}]$}}

\put(-85,80){\text{\tiny $k_2=k_1=0.1\, h\,{\rm Mpc}^{-1}$}}
\put(-145,70){\rotatebox[]{90}{\text{\tiny\tiny $B^{(0)}(k)\times10^{9}$}}}
\put(-135,70){\rotatebox[]{90}{\text{\tiny\tiny $[(h^{-1}\,{\rm Mpc})^{6}]$}}}
\put(-145,27){\rotatebox[]{90}{\text{\tiny\tiny $B^{(0)}_{\rm BigMD}(k)/$}}}
\put(-135,27){\rotatebox[]{90}{\text{\tiny\tiny $B^{(0)}_{\rm Patchy}(k)$}}}

\put(-122.7,55.5){\text{\tiny\tiny0.5}}
\put(-119,63.5){\text{\tiny\tiny1}}
\put(-122.7,72){\text{\tiny\tiny1.5}}
\put(-119,80){\text{\tiny\tiny2}}
\put(-122.7,89){\text{\tiny\tiny2.5}}

\put(-119,15.){\text{\tiny\tiny0}}
\put(-122.7,21.5){\text{\tiny\tiny0.5}}
\put(-119,28.3){\text{\tiny\tiny1}}
\put(-122.7,35.7){\text{\tiny\tiny1.5}}
\put(-119,43){\text{\tiny\tiny2}}

\put(-117,10){\text{\tiny\tiny0}}
\put(-93,10){\text{\tiny\tiny0.05}}
\put(-67,10){\text{\tiny\tiny0.10}}
\put(-41,10){\text{\tiny\tiny0.15}}
\put(-15,10){\text{\tiny\tiny0.20}}

\hspace{-0.2cm}
\includegraphics[width=4.5cm]{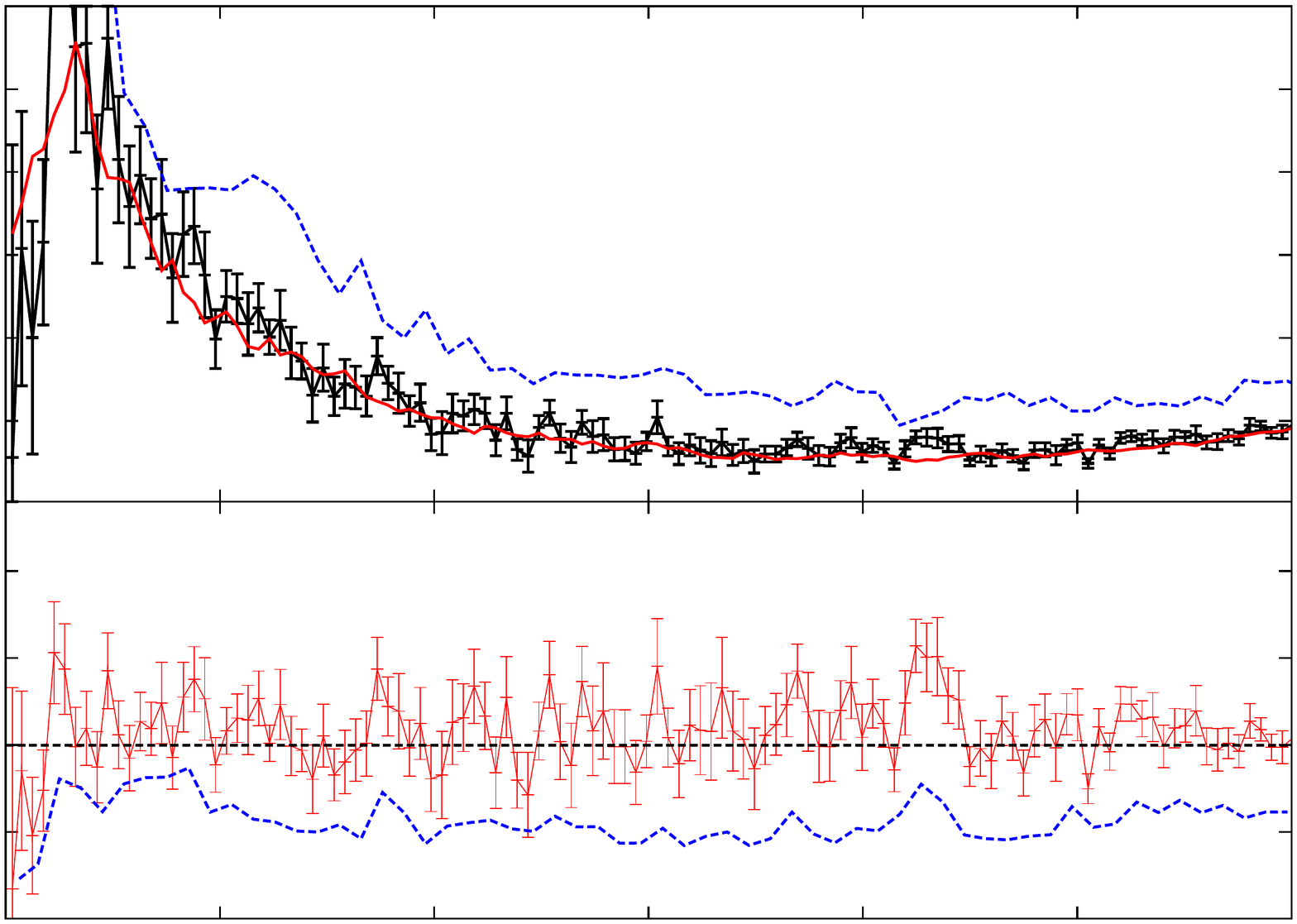}
\put(-80,2){\text{\tiny\tiny $k_3 \,[h\,{\rm Mpc}^{-1}]$}}

\put(-88,80){\text{\tiny $k_2=k_1=0.15\, h\,{\rm Mpc}^{-1}$}}

\put(-119,15.){\text{\tiny\tiny0}}
\put(-122.7,21.5){\text{\tiny\tiny0.5}}
\put(-119,28.3){\text{\tiny\tiny1}}
\put(-122.7,35.7){\text{\tiny\tiny1.5}}
\put(-119,43){\text{\tiny\tiny2}}

\put(-122.7,55.){\text{\tiny\tiny0.2}}
\put(-122.7,62.){\text{\tiny\tiny0.4}}
\put(-122.7,68.5){\text{\tiny\tiny0.6}}
\put(-122.7,75.5){\text{\tiny\tiny0.8}}
\put(-122.7,82.5){\text{\tiny\tiny1.0}}
\put(-122.7,89){\text{\tiny\tiny1.2}}

\put(-117,10){\text{\tiny\tiny0}}
\put(-102,10){\text{\tiny\tiny0.05}}
\put(-85,10){\text{\tiny\tiny0.10}}
\put(-67,10){\text{\tiny\tiny0.15}}
\put(-48,10){\text{\tiny\tiny0.20}}
\put(-32,10){\text{\tiny\tiny0.25}}
\put(-15,10){\text{\tiny\tiny0.30}}

\hspace{-0.2cm}
\includegraphics[width=4.5cm]{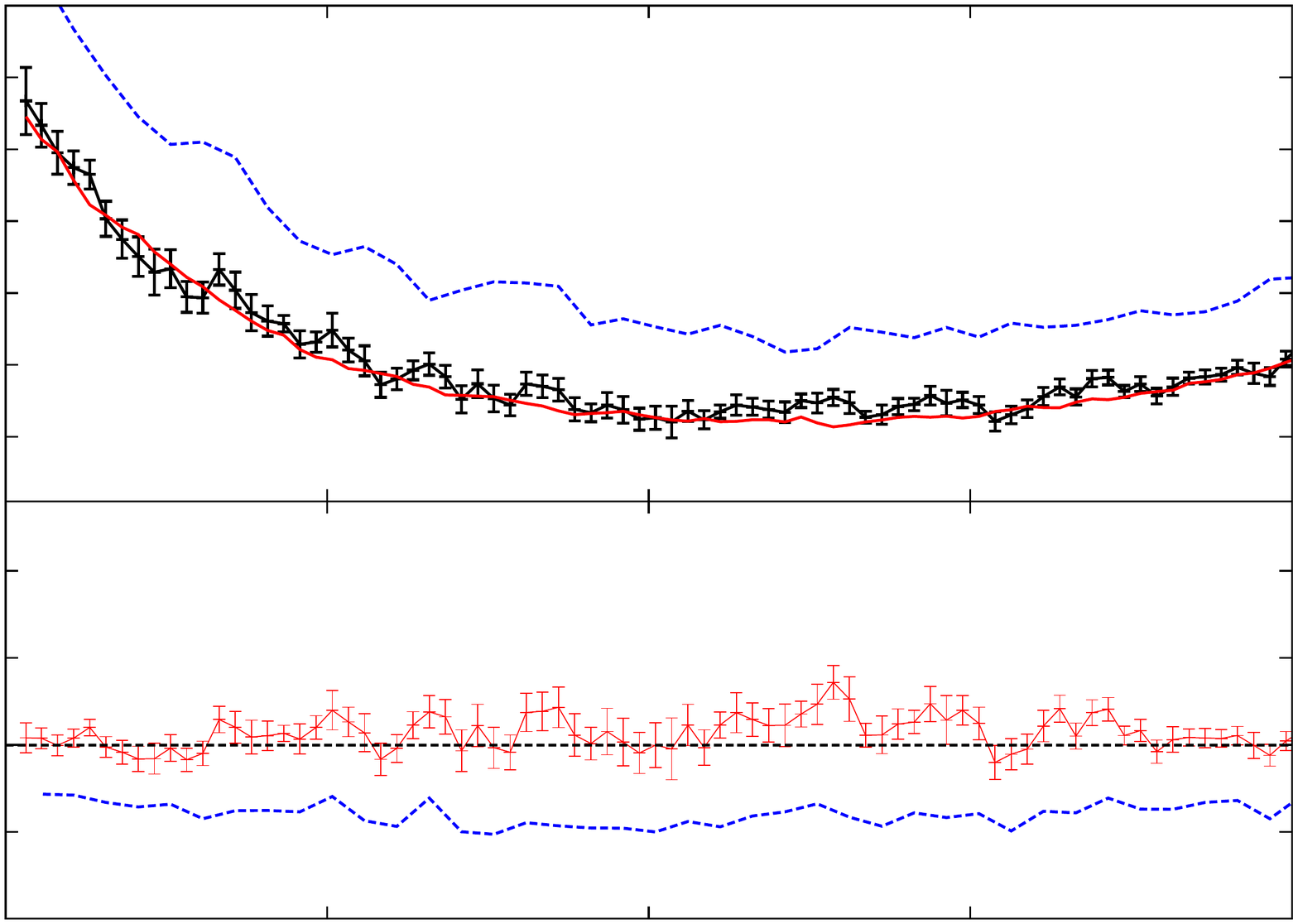}
\put(-80,2){\text{\tiny\tiny $k_3 \,[h\,{\rm Mpc}^{-1}]$}}

\put(-122.7,53.5){\text{\tiny\tiny0.1}}
\put(-122.7,59.8){\text{\tiny\tiny0.2}}
\put(-122.7,65.5){\text{\tiny\tiny0.3}}
\put(-122.7,71.5){\text{\tiny\tiny0.4}}
\put(-122.7,77.5){\text{\tiny\tiny0.5}}
\put(-122.7,83.5){\text{\tiny\tiny0.6}}
\put(-122.7,89){\text{\tiny\tiny0.7}}

\put(-119,15.){\text{\tiny\tiny0}}
\put(-122.7,21.5){\text{\tiny\tiny0.5}}
\put(-119,28.3){\text{\tiny\tiny1}}
\put(-122.7,35.7){\text{\tiny\tiny1.5}}
\put(-119,43){\text{\tiny\tiny2}}

\put(-118,10){\text{\tiny\tiny0.10}}
\put(-93,10){\text{\tiny\tiny0.15}}
\put(-67,10){\text{\tiny\tiny0.20}}
\put(-41,10){\text{\tiny\tiny0.25}}
\put(-15,10){\text{\tiny\tiny0.30}}

\put(-88,80){\text{\tiny $k_2=2k_1=0.2\, h\,{\rm Mpc}^{-1}$}}
\hspace{-0.2cm}
\includegraphics[width=4.5cm]{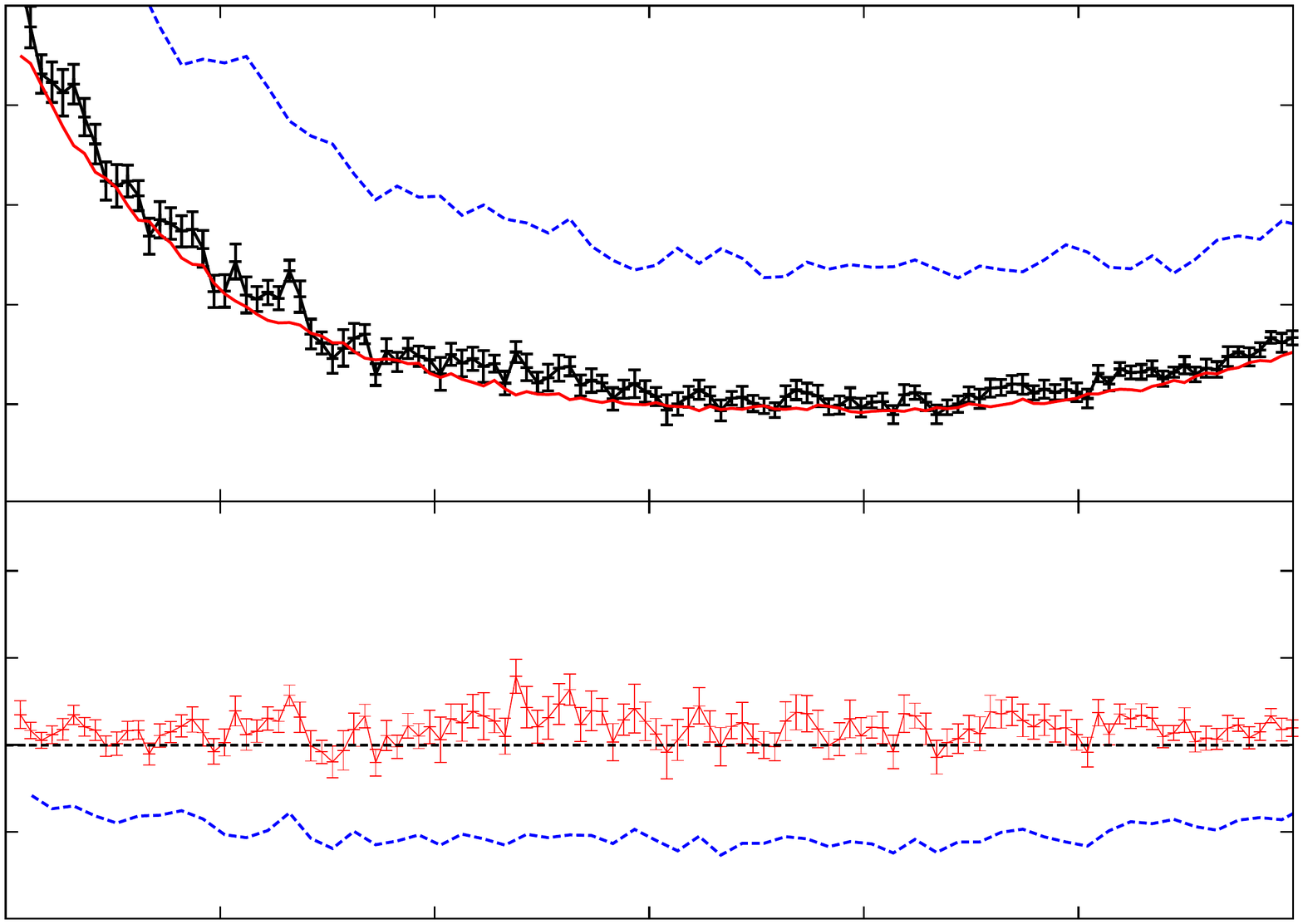}
\put(-80,2){\text{\tiny\tiny $k_3 \,[h\,{\rm Mpc}^{-1}]$}}

\put(-122.7,56.5){\text{\tiny\tiny0.4}}
\put(-122.7,65.){\text{\tiny\tiny0.8}}
\put(-122.7,73){\text{\tiny\tiny1.2}}
\put(-122.7,81.6){\text{\tiny\tiny1.6}}
\put(-119,89){\text{\tiny\tiny2}}

\put(-119,15.){\text{\tiny\tiny0}}
\put(-122.7,21.5){\text{\tiny\tiny0.5}}
\put(-119,28.3){\text{\tiny\tiny1}}
\put(-122.7,35.7){\text{\tiny\tiny1.5}}
\put(-119,43){\text{\tiny\tiny2}}

\put(-91,80){\text{\tiny $k_2=2k_1=0.3\, h\,{\rm Mpc}^{-1}$}}

\put(-117,10){\text{\tiny\tiny0}}
\put(-102,10){\text{\tiny\tiny0.05}}
\put(-85,10){\text{\tiny\tiny0.10}}
\put(-67,10){\text{\tiny\tiny0.15}}
\put(-48,10){\text{\tiny\tiny0.20}}
\put(-32,10){\text{\tiny\tiny0.25}}
\put(-15,10){\text{\tiny\tiny0.30}}

\end{tabular}
\caption{\label{fig:bik} Halo bispectrum for the \textsc{BigMD} $N$-body simulations and \textsc{patchy} mocks as a function of $k_3$. Top panels show the bispectrum in real space and bottom panels the bispectrum monopole in redshift space. Different columns present different scales and shapes as labeled. Same colour notation that in Fig.~\ref{fig:pk} is used.
}
\end{figure*}

\begin{figure*}
\begin{tabular}{cccc}

\includegraphics[width=4.5cm]{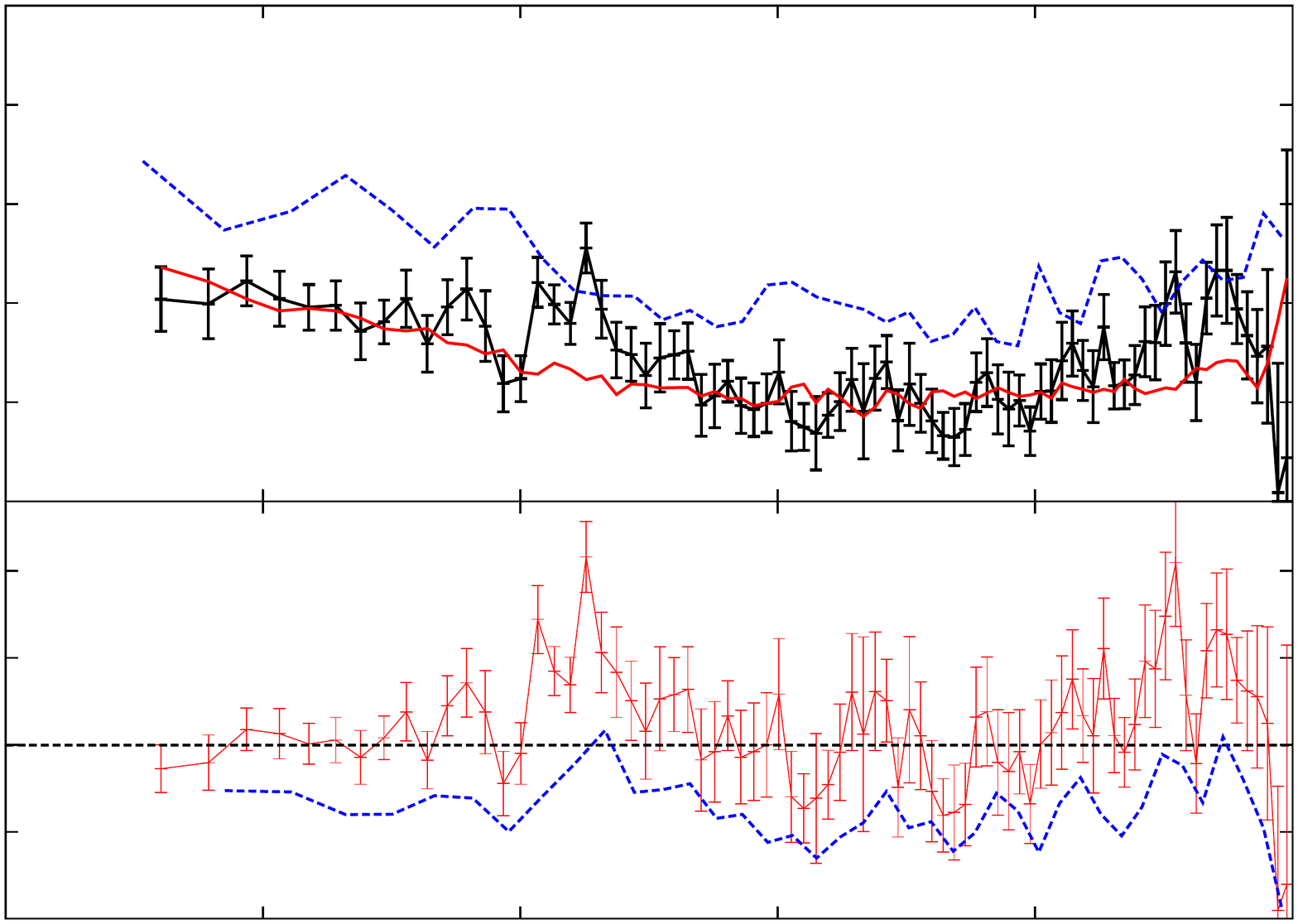}
\put(-84,80){\text{\tiny $k_2=k_1=0.1\, h\,{\rm Mpc}^{-1}$}}
\put(-135,70){\rotatebox[]{90}{\text{\tiny\tiny $Q(k)$}}}
\put(-145,27){\rotatebox[]{90}{\text{\tiny\tiny $Q_{\rm BigMD}(k)/$}}}
\put(-135,27){\rotatebox[]{90}{\text{\tiny\tiny $Q_{\rm Patchy}(k)$}}}

\put(-122.7,56.5){\text{\tiny\tiny0.4}}
\put(-122.7,65.){\text{\tiny\tiny0.8}}
\put(-122.7,73){\text{\tiny\tiny1.2}}
\put(-122.7,81.6){\text{\tiny\tiny1.6}}
\put(-119,89){\text{\tiny\tiny2}}


\put(-119,15.){\text{\tiny\tiny0}}
\put(-122.7,21.5){\text{\tiny\tiny0.5}}
\put(-119,28.3){\text{\tiny\tiny1}}
\put(-122.7,35.7){\text{\tiny\tiny1.5}}
\put(-119,43){\text{\tiny\tiny2}}

\hspace{-0.2cm}
\includegraphics[width=4.5cm]{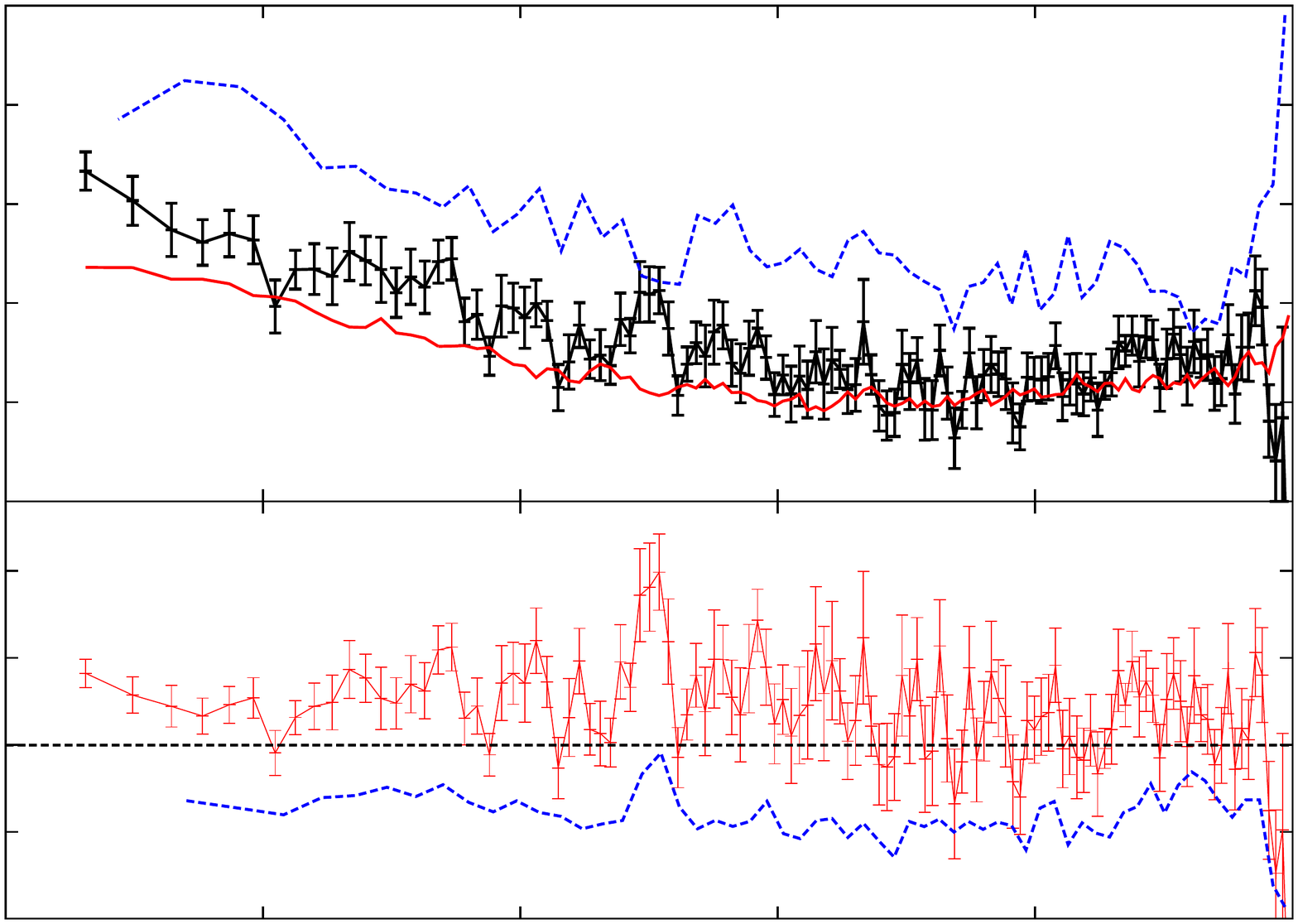}
\put(-90,80){\text{\tiny $k_2=k_1=0.15\, h\,{\rm Mpc}^{-1}$}}

\put(-122.7,56.5){\text{\tiny\tiny0.4}}
\put(-122.7,65.){\text{\tiny\tiny0.8}}
\put(-122.7,73){\text{\tiny\tiny1.2}}
\put(-122.7,81.6){\text{\tiny\tiny1.6}}
\put(-119,89){\text{\tiny\tiny2}}

\put(-119,15.){\text{\tiny\tiny0}}
\put(-122.7,21.5){\text{\tiny\tiny0.5}}
\put(-119,28.3){\text{\tiny\tiny1}}
\put(-122.7,35.7){\text{\tiny\tiny1.5}}
\put(-119,43){\text{\tiny\tiny2}}

\hspace{-0.2cm}
\includegraphics[width=4.5cm]{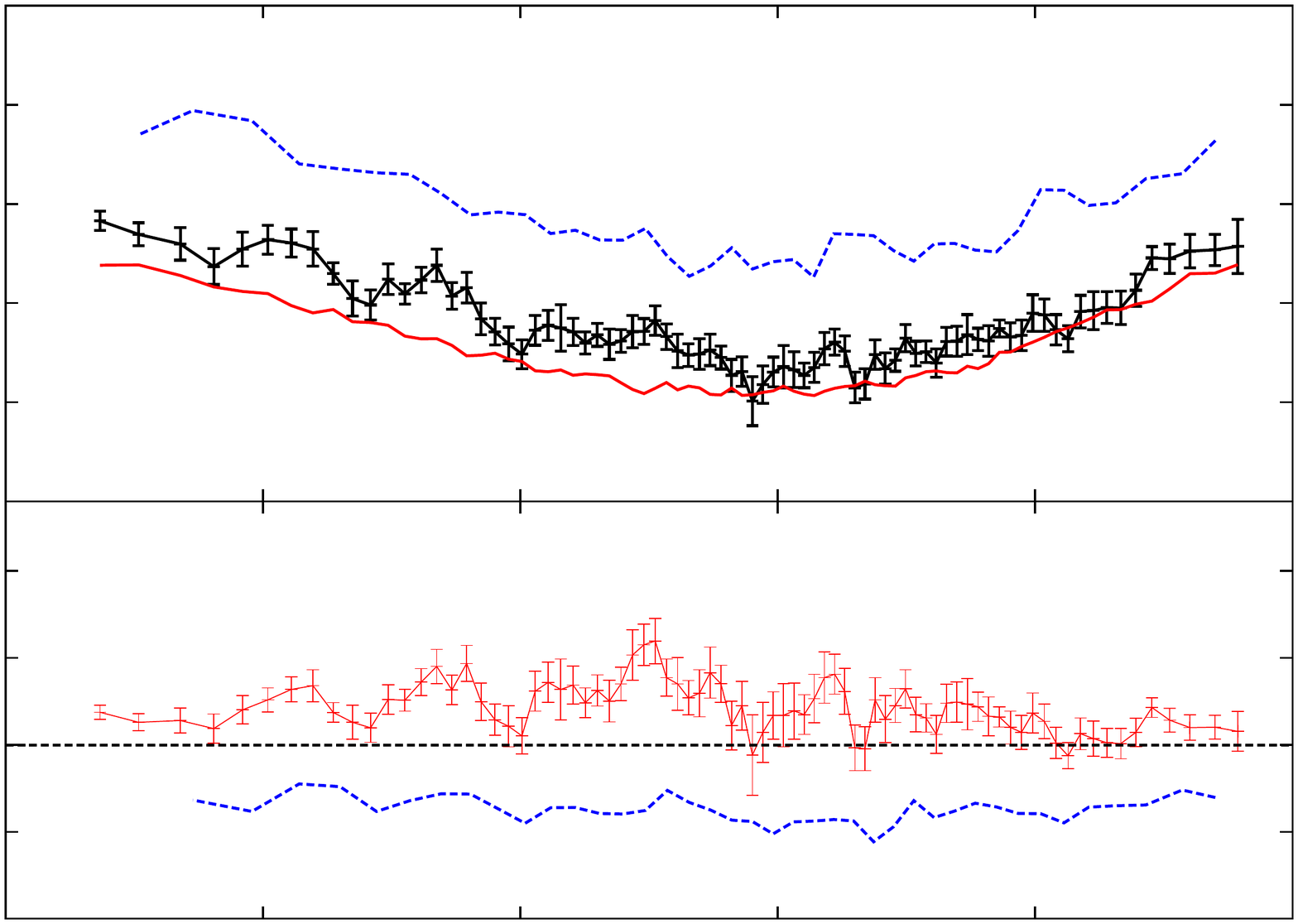}
\put(-90,80){\text{\tiny $k_2=2k_1=0.2\, h\,{\rm Mpc}^{-1}$}}

\put(-122.7,56.5){\text{\tiny\tiny0.4}}
\put(-122.7,65.){\text{\tiny\tiny0.8}}
\put(-122.7,73){\text{\tiny\tiny1.2}}
\put(-122.7,81.6){\text{\tiny\tiny1.6}}
\put(-119,89){\text{\tiny\tiny2}}

\put(-119,15.){\text{\tiny\tiny0}}
\put(-122.7,21.5){\text{\tiny\tiny0.5}}
\put(-119,28.3){\text{\tiny\tiny1}}
\put(-122.7,35.7){\text{\tiny\tiny1.5}}
\put(-119,43){\text{\tiny\tiny2}}

\hspace{-0.2cm}
\includegraphics[width=4.5cm]{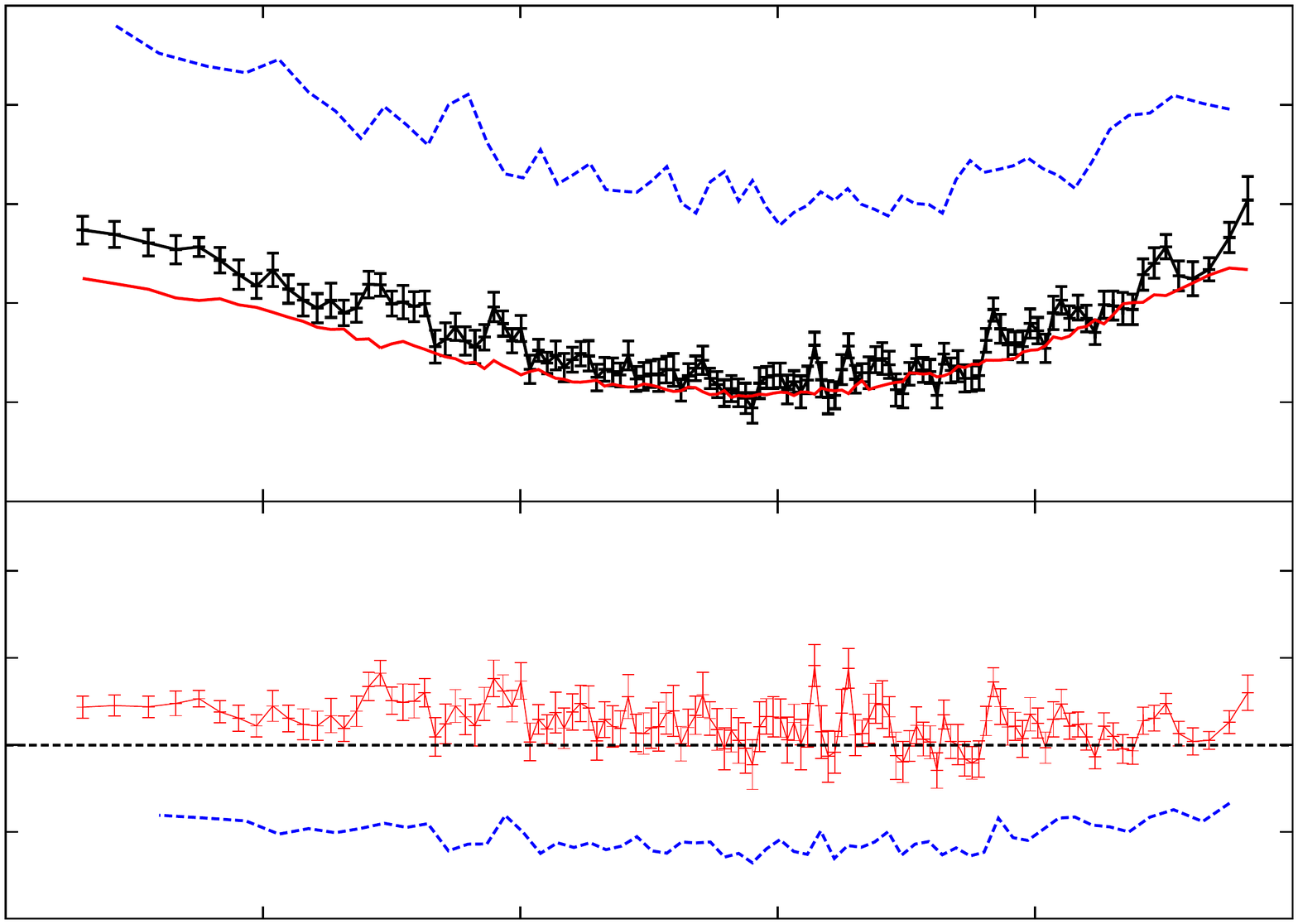}
\put(-90,80){\text{\tiny $k_2=2k_1=0.3\, h\,{\rm Mpc}^{-1}$}}

\put(-122.7,56.5){\text{\tiny\tiny0.4}}
\put(-122.7,65.){\text{\tiny\tiny0.8}}
\put(-122.7,73){\text{\tiny\tiny1.2}}
\put(-122.7,81.6){\text{\tiny\tiny1.6}}
\put(-119,89){\text{\tiny\tiny2}}

\put(-119,15.){\text{\tiny\tiny0}}
\put(-122.7,21.5){\text{\tiny\tiny0.5}}
\put(-119,28.3){\text{\tiny\tiny1}}
\put(-122.7,35.7){\text{\tiny\tiny1.5}}
\put(-119,43){\text{\tiny\tiny2}}

\vspace{-.5cm}
\\

\includegraphics[width=4.5cm]{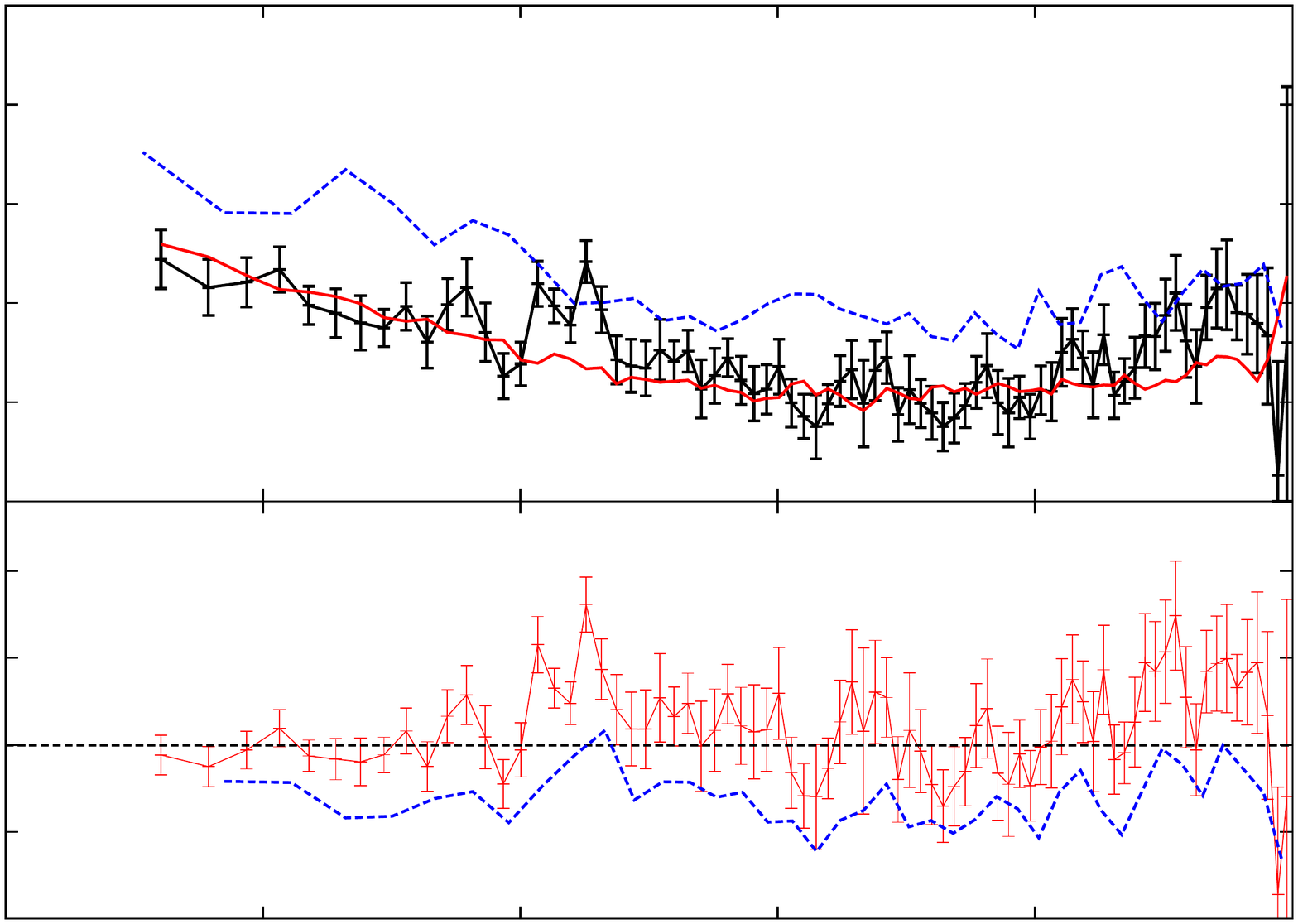}
\put(-70,2){\text{\tiny\tiny $\alpha_{12}/\pi$}}

\put(-84,80){\text{\tiny $k_2=k_1=0.1\, h\,{\rm Mpc}^{-1}$}}
\put(-135,70){\rotatebox[]{90}{\text{\tiny\tiny $Q^{(0)}(k)$}}}
\put(-145,27){\rotatebox[]{90}{\text{\tiny\tiny $Q^{(0)}_{\rm BigMD}(k)/$}}}
\put(-135,27){\rotatebox[]{90}{\text{\tiny\tiny $Q^{(0)}_{\rm Patchy}(k)$}}}

\put(-122.7,56.5){\text{\tiny\tiny0.4}}
\put(-122.7,65.){\text{\tiny\tiny0.8}}
\put(-122.7,73){\text{\tiny\tiny1.2}}
\put(-122.7,81.6){\text{\tiny\tiny1.6}}
\put(-119,89){\text{\tiny\tiny2}}

\put(-119,15.){\text{\tiny\tiny0}}
\put(-122.7,21.5){\text{\tiny\tiny0.5}}
\put(-119,28.3){\text{\tiny\tiny1}}
\put(-122.7,35.7){\text{\tiny\tiny1.5}}
\put(-119,43){\text{\tiny\tiny2}}


\put(-117,10){\text{\tiny\tiny0}}
\put(-96,10){\text{\tiny\tiny0.2}}
\put(-75,10){\text{\tiny\tiny0.4}}
\put(-55,10){\text{\tiny\tiny0.6}}
\put(-33,10){\text{\tiny\tiny0.8}}
\put(-11,10){\text{\tiny\tiny1}}

\hspace{-0.2cm}
\includegraphics[width=4.5cm]{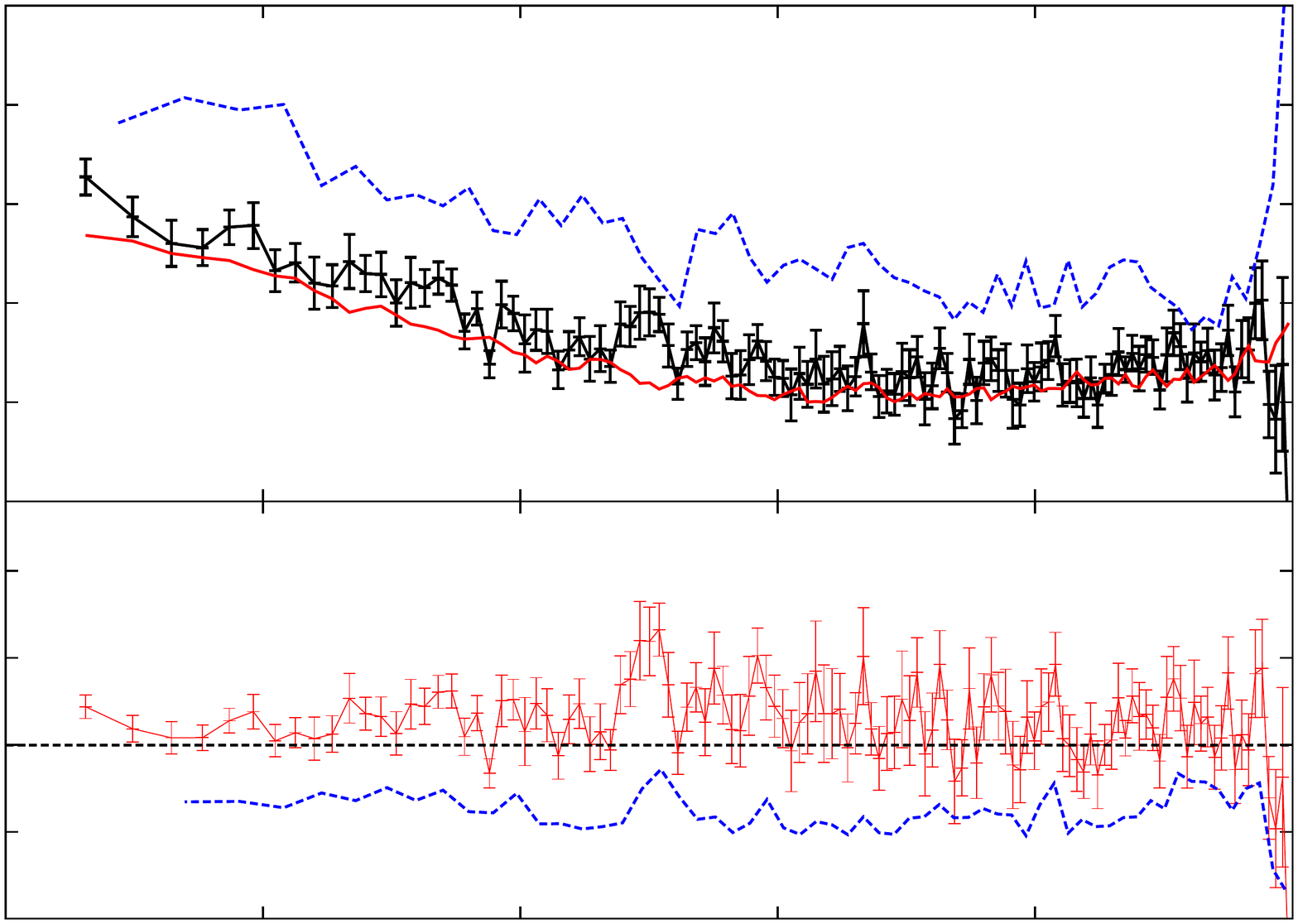}
\put(-70,2){\text{\tiny\tiny $\alpha_{12}/\pi$}}

\put(-90,80){\text{\tiny $k_2=k_1=0.15\, h\,{\rm Mpc}^{-1}$}}

\put(-122.7,56.5){\text{\tiny\tiny0.4}}
\put(-122.7,65.){\text{\tiny\tiny0.8}}
\put(-122.7,73){\text{\tiny\tiny1.2}}
\put(-122.7,81.6){\text{\tiny\tiny1.6}}
\put(-119,89){\text{\tiny\tiny2}}

\put(-119,15.){\text{\tiny\tiny0}}
\put(-122.7,21.5){\text{\tiny\tiny0.5}}
\put(-119,28.3){\text{\tiny\tiny1}}
\put(-122.7,35.7){\text{\tiny\tiny1.5}}
\put(-119,43){\text{\tiny\tiny2}}

\put(-117,10){\text{\tiny\tiny0}}
\put(-96,10){\text{\tiny\tiny0.2}}
\put(-75,10){\text{\tiny\tiny0.4}}
\put(-55,10){\text{\tiny\tiny0.6}}
\put(-33,10){\text{\tiny\tiny0.8}}
\put(-11,10){\text{\tiny\tiny1}}

\hspace{-0.2cm}
\includegraphics[width=4.5cm]{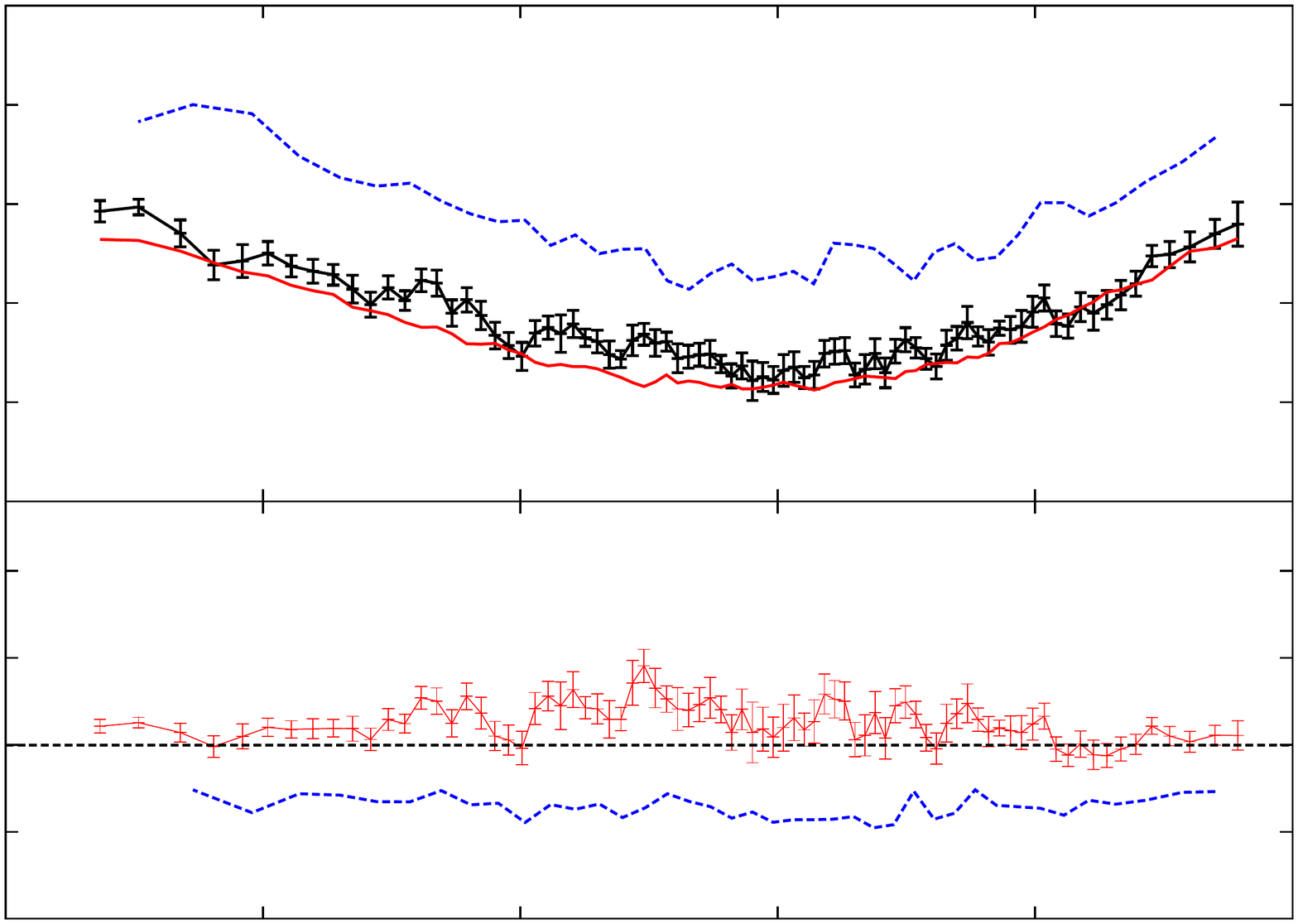}
\put(-70,2){\text{\tiny\tiny $\alpha_{12}/\pi$}}

\put(-90,80){\text{\tiny $k_2=2k_1=0.2\, h\,{\rm Mpc}^{-1}$}}

\put(-122.7,56.5){\text{\tiny\tiny0.4}}
\put(-122.7,65.){\text{\tiny\tiny0.8}}
\put(-122.7,73){\text{\tiny\tiny1.2}}
\put(-122.7,81.6){\text{\tiny\tiny1.6}}
\put(-119,89){\text{\tiny\tiny2}}

\put(-119,15.){\text{\tiny\tiny0}}
\put(-122.7,21.5){\text{\tiny\tiny0.5}}
\put(-119,28.3){\text{\tiny\tiny1}}
\put(-122.7,35.7){\text{\tiny\tiny1.5}}
\put(-119,43){\text{\tiny\tiny2}}

\put(-117,10){\text{\tiny\tiny0}}
\put(-96,10){\text{\tiny\tiny0.2}}
\put(-75,10){\text{\tiny\tiny0.4}}
\put(-55,10){\text{\tiny\tiny0.6}}
\put(-33,10){\text{\tiny\tiny0.8}}
\put(-11,10){\text{\tiny\tiny1}}

\hspace{-0.2cm}
\includegraphics[width=4.5cm]{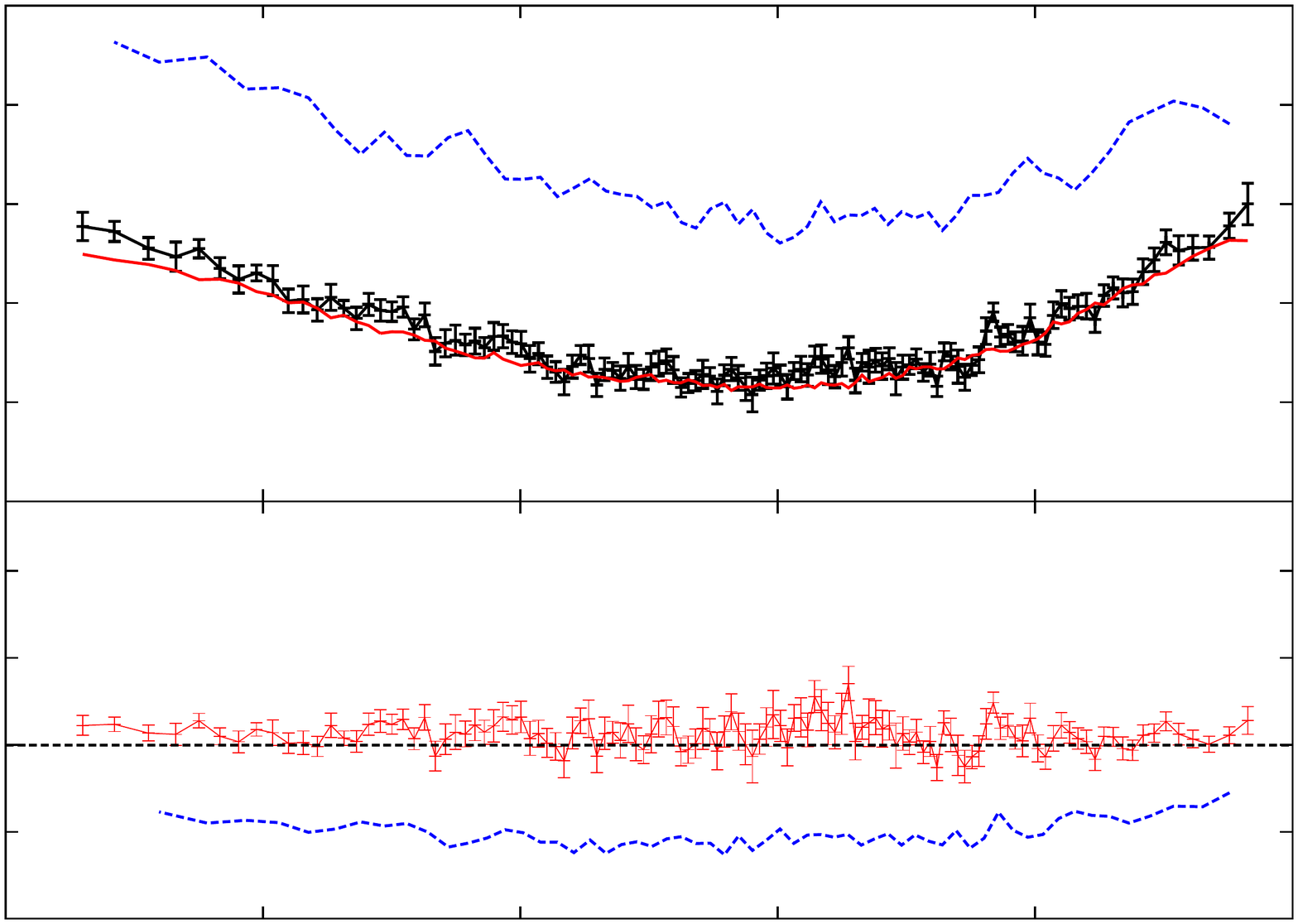}
\put(-70,2){\text{\tiny\tiny $\alpha_{12}/\pi$}}

\put(-90,80){\text{\tiny $k_2=2k_1=0.3\, h\,{\rm Mpc}^{-1}$}}

\put(-122.7,56.5){\text{\tiny\tiny0.4}}
\put(-122.7,65.){\text{\tiny\tiny0.8}}
\put(-122.7,73){\text{\tiny\tiny1.2}}
\put(-122.7,81.6){\text{\tiny\tiny1.6}}
\put(-119,89){\text{\tiny\tiny2}}

\put(-119,15.){\text{\tiny\tiny0}}
\put(-122.7,21.5){\text{\tiny\tiny0.5}}
\put(-119,28.3){\text{\tiny\tiny1}}
\put(-122.7,35.7){\text{\tiny\tiny1.5}}
\put(-119,43){\text{\tiny\tiny2}}

\put(-117,10){\text{\tiny\tiny0}}
\put(-96,10){\text{\tiny\tiny0.2}}
\put(-75,10){\text{\tiny\tiny0.4}}
\put(-55,10){\text{\tiny\tiny0.6}}
\put(-33,10){\text{\tiny\tiny0.8}}
\put(-11,10){\text{\tiny\tiny1}}

\end{tabular}
\caption{\label{fig:qk} Halo reduced bispectrum for \textsc{BigMD} $N$-body simulations and \textsc{patchy} mocks as a function of the angle between ${\bf k}_1$ and ${\bf k}_2$, $\alpha_{12}$.  Same colour and panel notation that in Fig.~\ref{fig:bik} is used.}
\end{figure*}

The measurement of the bispectrum is performed in a similar way of the approach described in \citet[][]{Gil-Marin_et_al_2012}. This method consists of generating  $k$-triangles and randomly-orientate them in $k$-space. When the number of random triangles is sufficiently large, the mean value of their bispectra tends to the fiducial bispectrum \citep[for details see][]{Gil-Marin_et_al_2012}.

Discreteness adds a spurious contribution to the measured power spectrum and bispectrum. In this paper we assume that these contributions are of Poisson type and therefore are given by,
\begin{eqnarray}
P_{\rm sn}(k)&=&\frac{1}{\bar n} \\
B_{\rm sn}({\bf k}_1,{\bf k}_2)&=&\frac{1}{\bar n}\left[ P(k_1)+P(k_2)+P(k_3)\right]+\frac{1}{\bar{n}^2}
\end{eqnarray}
where $k_3=|{\bf k}_1+{\bf k}_2|$ and $\bar{n}$ is the number density of haloes.
We are aware that deviations from Poissonity are present (see \S \ref{sec:statsprob}). However, for the purposes of this paper, we find it sufficient to use the Poisson predictions to correct the power spectrum and bispectrum measurements in a consistent way for both the $N$-body and the \textsc{patchy} simulations.

Both for the power spectrum and bispectrum the errors associated to the measurement come from the dispersion of 20 independent realizations of the \textsc{patchy} mocks.

Recall that the \textsc{BigMD} simulation corresponds to a single  box of $2.5\,h^{-1}{\rm Gpc}$ on a side, and hence, has an effective volume of $V_{\rm eff}=15.625\,[h^{-1}{\rm Gpc}]^3$. On the other hand, for the \textsc{patchy} mocks simulations we dispose of 20 boxes of $2.5\,h^{-1}{\rm Gpc}$ on a side with independent initial conditions and with a total effective volume of $V_{\rm eff}=312.5\,[h^{-1}{\rm Gpc}]^3$. Due to this difference in effective volumes, \textsc{BigMD} measurements (for both power spectrum and bispectrum)  present a more noisy behaviour than \textsc{patchy} measurements, with error-bars that are $\sim\sqrt{20}$ times larger for \textsc{BigMD} compared to \textsc{patchy}. This difference is more evident for the bispectrum  because the signal-to-noise is weaker, and therefore, the errors larger than for the power spectrum. 

For both power spectrum and bispectrum we present the \textsc{BigMD} error-bars computed from the dispersion among the 20 realizations of \textsc{patchy} mocks. We do not display any error-bars for the \textsc{patchy} mocks measurements for clarity. These error-bars would be $\sqrt{20}$ times smaller than the ones showed for \textsc{BigMD}.

Fig.~\ref{fig:pk} presents the comparison between the power spectrum of \textsc{BigMD} $N$-body simulations  (black symbols), the \textsc{patchy} mocks with $\rho^{\rm High}$ (red lines), and the single \textsc{patchy} mock with $\rho^{\rm Low}$ (dashed-blue line). The left panel displays the real space power spectrum and the right panel the redshift space power spectrum monopole. Bottom sub-panels present the relative deviation between them. As a  general trend, we see a good agreement between \textsc{BigMD} and \textsc{patchy} power-spectra, which agree within $\leq2\%$ accuracy for $k\leq0.35\,h\,{\rm Mpc}^{-1}$, both in real and redshift space. We observe that at small scales, the \textsc{patchy} mocks considered in this study tend to slightly under-predict the power spectrum by $\sim2\%$ respect to \textsc{BigMD}, being compatible within better than  10 \% up to $k=1\,h\,{\rm Mpc}^{-1}$. Further tests have, however, shown that a different set of parameters can also over-predict the power spectrum at large scales, indicating that some set will display a better fit on small scales. We do not intend in this study to reach precisions of 2\% beyond the scale relevant to BAOs ($k=0.35\,h\,{\rm Mpc}^{-1}$) and therefore have not further improved the parameters. 
The \textsc{patchy} mocks with  $\rho_{\rm th}^{\rm High}$  slightly over-predict the power spectrum towards high $k$, however remaining compatible within 2\% with the \textsc{BigMD} simulation. The non-linear contribution of redshift space distortions responsible of damping the power spectrum on small scales can be controlled with the  factor of the dispersion term as shown in \citet[][]{patchy}. We plan to investigate this further including calculations of the quadrupole in a future work.

Fig.~\ref{fig:bik} presents the comparison between \textsc{BigMD} and \textsc{patchy} bispectrum using the same colour notation that in Fig.~\ref{fig:pk}.  Top panels correspond to real space bispectrum and bottom panels to redshift space bispectrum monopole. Different columns correspond to different scales and shapes as indicated. Note that in this case, the precision of \textsc{BigMD} measurements is a limiting factor when we test the accuracy of \textsc{patchy} respect to \textsc{BigMD} bispectra. As a general trend we see that both \textsc{patchy} and \textsc{BigMD} agree within $10-20\%$ accuracy. Similarly to the power spectrum case, we see that \textsc{patchy} tends to under-predict the bispectrum at small scales by $\sim10-20\%$ with no evidence of any shape dependence.

Fig.~\ref{fig:qk} presents a similar comparison for the reduced bispectra using the same colour and panel notation that in Fig.~\ref{fig:pk} and \ref{fig:bik}.
Similarly to the bispectrum case, we observe a general agreement between \textsc{BigMD} and \textsc{patchy} with $\rho^{\rm High}$ reduced bispectra within $10-20\%$ accuracy. However, we see that the \textsc{patchy}  with $\rho^{\rm High}$ prediction for the reduced bispectra tends to under-estimate the \textsc{BigMD} prediction by $\sim20\%$. This deviation tends to be more evident in real space that in redshift space. However, we should note the large uncertainties in our single $N$-body simulation shown in the fluctuations as a function of the angle $\alpha_{12}$. 

We have tested that a different set of bias parameters matching the power spectrum, but disregarding the shape of the halo PDF produces bispectra, which can deviate from the true one by about  a factor of 2 as can be seen represented by the dashed-blue curves  corresponding to \textsc{patchy} with $\rho^{\rm Low}$ in Figs.~\ref{fig:pk}, \ref{fig:bik} and \ref{fig:qk}.

In summary, we consider that the bispectra of the \textsc{patchy}-mocks fit well the ones from \textsc{BigMD} given the uncertainties of our single reference $N$-body simulation. Further investigation should be done in the future with a larger number of reference $N$-body simulations.

\section{Discussion and conclusions}

\label{sec:conc}

In this work we have presented a method  to produce mock galaxy catalogues with efficient perturbation theory schemes, which match the number density, power spectra and bispectra in real and in redshift space from $N$-body simulations.
The ingredients of our scheme are given by  an improved Lagrangian perturbation theory based approach to describe dark matter structure formation and a non-linear stochastic bias model \citep[the \textsc{patchy}-code:][]{patchy}. The essential contribution of this work is the way in which we constrain the parameters of our bias model. In addition of aiming at  reproducing the two-point statistics  we need to additionally constrain the univariate halo probability distribution function encoding higher-order correlation functions. We have shown that this approach permits us to reasonably reproduce the bispectrum. Nevertheless, we have not included any explicit constraint from the three-point statistics. We therefore expect that our general approach yields also reasonable fits of the four-point correlations, which are relevant, as they quantify the sample variance and covariance of  two-point statistics measurements \citep[][]{Cooray01,Takada2013}. We leave a thourough analysis of covariance matrices for a forthcoming publication.

We have used a Luminous Red Galaxy (LRG) like  reference halo catalogue  with the typical BOSS CMASS galaxy number density at $z=0.577$ based on one of the \textsc{BigMD} $N$-body simulations.
We have demonstrated that halo catalogues based on the same underlying dark matter field with a fix halo number density (i.e., matching the integral of the halo PDF) and accurately matching the power spectrum (within 2\% for $k\leq0.35\,h\,{\rm Mpc}^{-1}$ and deviating less than 10\% up to $k=1\,h\,{\rm Mpc}^{-1}$),  can lead to very different bispectra depending on the halo bias model. A model ignoring the shape of the halo PDF can lead to deviations up to the level of a factor of 2.  The catalogues obtained additionally constraining the halo PDF can significantly lower the discrepancy in the three-point statistics yielding closely unbiased bispectra both in real and in redshift space,  which are compatible with those corresponding to an $N$-body simulation in general within 10\% (deviating at most up to 20\%).

Our calculations have shown that the constant linear bias of $\sim2$ for  LRG-like galaxies found in the power spectrum (defined as the square root of the halo power spectrum divided by the dark matter power spectrum), mainly comes from sampling halos in the high density peaks choosing a high density threshold rather than from a factor multiplying the dark matter density field.

The method presented in this work can be applied to directly produce galaxy catalogues without requiring the halo distribution, as it just cares about the statistical properties for the type of tracers of interest,  and is in this way a model-independent method. The model dependency comes in, when the method is calibrated with a particular mock galaxy catalogue. This dependence could be broken, by using as the reference an observed sample of galaxies.  We could then add the large-scale modes and produce different phases with the statistics constrained by the observations with our \textsc{patchy}  approach.
In a separated work we will show how to use a mock galaxy sample based on abundance matching to produce mocks with \textsc{patchy}  (Scoccola et al., in prep.).

We plan to address in forthcoming works a number of studies which have not been done here, like investigating the impact of (a deterministic) non-local bias, computing the two-point and three-point correlation functions in configuration space and further improving the halo PDF fit. 

Our method contributes towards an efficient modelling of the halo/galaxy distribution required to estimate uncertainties in our measurements from galaxy redshift surveys. We have also demonstrated that it represents a powerful tool to test various bias models.

%

\vspace{0.1cm}
\begin{flushleft}
{\bf Acknowledgments}
\end{flushleft}
The authors thank Mark Neyrinck and Shun Saito for encouraging discussions. HGM is grateful for support from the UK Science and Technology Facilities Council through the grant ST/I001204/1. CGS acknowledges funding from the Spanish Ministry of Economy and Competitiveness (MINECO) under the project AYA2012-3972-C02-01. CC and FP were supported by the Spanish MICINN’s Consolider-Ingenio 2010 Programme under grant MultiDark CSD2009-00064 and AYA2010-21231-C02-01 grant, the Comunidad de Madrid under grant HEPHACOS S2009/ESP-1473, and Spanish MINECO’s “Centro de Excelencia Severo Ochoa” Programme under grant SEV-2012-0249. GY and FP acknowledge support from the Spanish MINECO under research grants  AYA2012-31101, FPA2012-34694, AYA2010-21231, Consolider Ingenio SyeC CSD2007-0050 and  from Comunidad de Madrid under  ASTROMADRID  project (S2009/ESP-1496). 
 Power spectrum and bispectrum computations were performed on the Sciama High Performance Compute (HPC) cluster which is supported by the ICG, SEPNet.  The MultiDark Database used in this paper and the web application providing online access to it were constructed as part of the activities of the German Astrophysical Virtual Observatory as result of a collaboration between the Leibniz-Institute for Astrophysics Potsdam (AIP) and the Spanish MultiDark Consolider Project CSD2009-00064.  The  \textsc{BigMD} simulation suite have been performed in the Supermuc supercomputer at LRZ using time granted by PRACE.
\vspace{-.6cm}

{\small
\bibliographystyle{mn2e}
\bibliography{lit}
}

\end{document}